\documentclass[preprint2]{proto}
\usepackage{times}

\newcommand{\refs}{\par\noindent\hangindent=1pc\hangafter=1}
\def\pasp{{\em Publ. Astron. Soc. Pac.}}
\def\pasj{{\em Publ. Astron. Soc. Japan}}

\def\tp{$T_{{\rm eff},pl}\,$}

\def\mic{$\mu$m$\,$}

\def\sles{\lower2pt\hbox{$\buildrel {\scriptstyle <}
   \over {\scriptstyle\sim}$}}
\def\sgreat{\lower2pt\hbox{$\buildrel {\scriptstyle >}
   \over {\scriptstyle\sim}$}}
\def\S{Section~}

\begin{document}

\title{\textbf{\LARGE When Extrasolar Planets Transit Their Parent Stars}}

\author{\textbf{David Charbonneau}}
\affil{\small\em Harvard-Smithsonian Center for Astrophysics}
\author{\textbf{Timothy M. Brown}}
\affil{\small\em High Altitude Observatory}
\author{\textbf{Adam Burrows}}
\affil{\small\em University of Arizona}
\author{\textbf{Greg Laughlin}}
\affil{\small\em University of California, Santa Cruz}

\begin{abstract}
\baselineskip = 11pt
\leftskip = 0.65in
\rightskip = 0.65in
\parindent=1pc
{\small 
When extrasolar planets are observed to transit their parent
stars, we are granted unprecedented access to their physical
properties.  It is only for transiting planets that we
are permitted direct estimates of the planetary masses
and radii, which provide the fundamental constraints on
models of their physical structure.  In particular, precise
determination of the radius may indicate the presence (or absence)
of a core of solid material, which in turn would speak to the
canonical formation model of gas accretion onto a core of
ice and rock embedded in a protoplanetary disk.  Furthermore,
the radii of planets in close proximity to their stars are
affected by tidal effects and the intense stellar radiation.
As a result, some of these ``hot Jupiters" are significantly
larger than Jupiter in radius.  Precision follow-up studies of
such objects (notably with the space-based platforms of the {\em Hubble}
and {\em Spitzer Space Telescopes}) have enabled
direct observation of their transmission spectra and emitted radiation.
These data provide the first observational constraints on atmospheric
models of these extrasolar gas giants, and permit a direct comparison
with the gas giants of the Solar system.  Despite significant
observational challenges, numerous transit surveys and quick-look
radial velocity surveys are active,
and promise to deliver an ever-increasing number of these precious
objects.  The detection of transits of short-period Neptune-sized objects, 
whose existence was recently uncovered by the radial-velocity surveys,
is eagerly anticipated.  Ultra-precise 
photometry enabled by upcoming space missions offers the
prospect of the first detection of an extrasolar Earth-like
planet in the habitable zone of its parent star,
just in time for Protostars and Planets VI.
 \\~\\~\\~}
 
\end{abstract}  

\section{\textbf{OVERVIEW}}
The month of October 2005, in which the fifth Protostars and Planets
meeting was held, marked two important events in the brief history
of the observational study of planets orbiting nearby, Sun-like stars.
First, it was the ten-year anniversary of the discovery of 51~Pegb
({\em Mayor and Queloz}, 1995), whose small orbital separation implied
that similar hot Jupiters could be found in orbits nearly co-planar
to our line of sight, resulting in mutual eclipses of the planet and
star.  Second, October 2005 heralded the discovery of the ninth such transiting
planet ({\em Bouchy et al.}, 2005a).  This select group of extrasolar planets
has enormous influence on our overall understanding of 
these objects:  The 9 transiting planets are the only ones for which we have accurate
estimates of key physical parameters such as mass, radius, and, by
inference, composition.  Furthermore, precise monitoring of these systems during primary and secondary
eclipse has permitted the direct study of their atmospheres.  As
a result, transiting planets are the only ones whose physical structure and 
atmospheres may be compared in detail to the planets of the Solar system,
and indeed October 2005 was notable for being the month in which the number of objects
in the former category surpassed the latter.

Our review of this rapidly-evolving field of study proceeds as follows.  
In \S2, we consider the
physical structure of these objects, beginning with a summary of the
observations (\S2.1) before turning to their impact on our theoretical
understanding (\S2.2).  In \S3, we consider the atmospheres of these 
planets, by first summarizing the challenges to modeling such systems
(\S3.1), and subsequently reviewing the detections and upper 
limits, and the inferences they permit (\S3.2).  We end by considering 
the future prospects (\S4) for learning about rocky planets beyond
the Solar system through the detection and characterization of such objects
in transiting configurations.

\section{\textbf{PHYSICAL STRUCTURE}}
\subsection{Observations}
{\em 2.1.1.\ Introduction.} When a planet transits, we
can accurately measure the orbital inclination, $i$, allowing us 
to evaluate the planetary mass $M_{pl}$ directly from 
the minimum mass value $M_{pl} \sin{i}$ determined from radial-velocity
observations and an estimate of the stellar mass, $M_{\star}$.  The planetary radius, $R_{pl}$,
can be obtained by measuring the fraction of the parent star's light
that is occulted, provided a reasonable estimate of the stellar
radius, $R_{\star}$, is available.  With the mass and radius in hand,
we can estimate such critically interesting quantities as the
average density and surface gravity.
Hence, the information gleaned from the transiting planets allows us to 
attempt to unravel the structure and composition of the larger class of extrasolar
planets, to understand formation and evolution processes (including
orbital evolution), and to elucidate physical processes that may be important
in planetary systems generically.
Fig.~1 shows the mass-radius relation for the 9 known transiting planets,
with Jupiter and Saturn added for comparison.
It is fortunate that the present small
sample of objects spans a moderate range in mass and radius, and appears
to contain both a preponderance of planets whose structure is fairly well
described by theory, as well as a few oddities that challenge our
present knowledge.

\begin{figure*}
\begin{center}
\includegraphics[angle=0, width=5.2in]{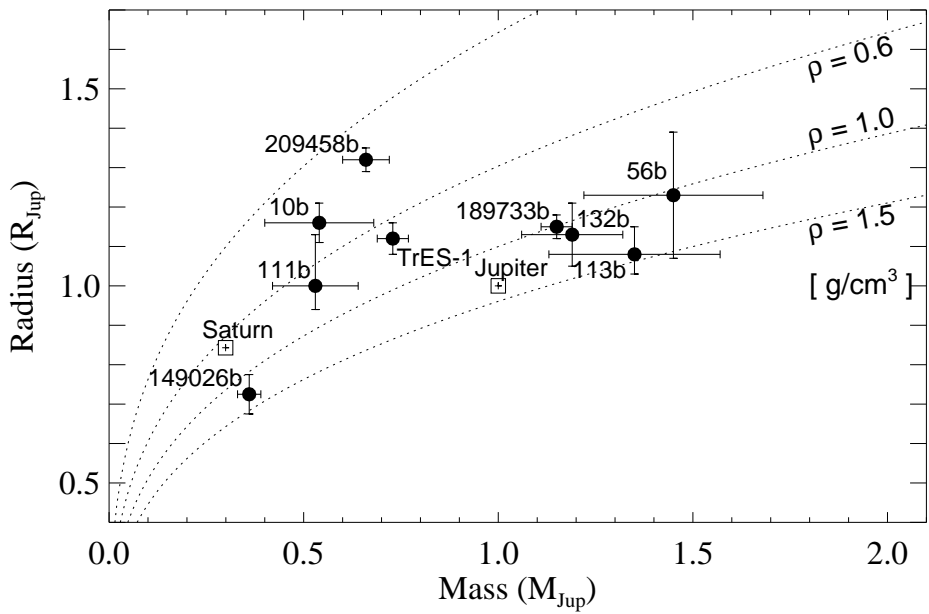}
\end{center}
\caption{\small Masses and radii for the 9 transiting planets,
as well as Jupiter and Saturn.  The data are tabulated in Table~1, and are
gathered from {\em Bakos et al., in preparation}, 
{\em Bouchy et al.} (2004, 2005b), {\em Brown et al., in preparation}, 
{\em Charbonneau et al.} (2006), {\em Holman et al.} (2005), {\em Knutson et al.} (2006),
{\em Laughlin et al.} (2005a), {\em Moutou et al.} (2004), {\em Pont et al.} (2004), {\em Sato et al.} (2006), 
{\em Sozzetti et al.} (2004), {\em Torres et al.} (2004a), and {\em Winn et al.} (2005).}
\end{figure*}

We begin by describing how the objects shown in Fig.~1 were
identified and characterized, and, along the way, we illuminate
the limitations that these methods imply for our efforts
to understand extrasolar planets as a class. 
By definition, transiting planets have their orbits oriented so that
the Earth lies nearly in their orbital plane.
This is an uncommon occurrence;  assuming random orientation of planetary
orbits, the probability that a planet with orbital eccentricity, $e$,
and longitude of periastron, $\varpi$, produces transits visible from the
Earth is given by
$$
P_{\rm tr}=0.0045 \left( {1 {\rm AU}}\over{a} \right) 
\left( {{R_{\star}+R_{pl}}\over{R_\odot}} \right)
\left[ {{1+e\cos(\frac{\pi}{2} - \varpi)} \over{1-e^2}}\right]
$$
which is inversely proportion to $a$, the orbital semi-major axis.
All known transiting planets have orbital eccentricities consistent
with zero, for which the last factor in the above equation reduces to unity.

The radii of Jovian planets are typically only about 10\% 
of the stellar radii.
The transits known to date result in a $0.3 - 3 \%$
diminution of the stellar flux reaching the Earth. 
These transits last for $1.5 - 3.5$ hours, and accurate ground-based
characterizations of these events are challenging.
The paucity and subtlety of the transits make it necessary to
use great care to reduce the random errors and systematic biases that
plague the estimation of the planets' fundamental properties (\S2.1.4).
\\

{\em 2.1.2.\ Methods of Detection.} The presently-known 
transiting planets have all been detected by one of 
the two following means, both foreseen by {\em Struve} (1952): (1)
Photometric detection of transit-like events, with subsequent confirmation
of planetary status via radial-velocity measurements, and (2)
radial-velocity detection of a planet with subsequent measurement
of photometric transits.
Radial velocity detection has the advantage that the planetary nature of the target
object is generally unambiguous.
Its disadvantage is that it requires substantial observing time on 
large telescopes to identify each planetary system, and only then can the relatively
cheap process of searching for photometric transits begin.
Direct photometric transit searches simultaneously monitor large numbers of stars
in a given field of view, but suffer from a very high rate of astrophysical false
positives (\S2.1.3).

Successful photometric transit searches have so far adopted one of two basic strategies,
using either moderate-sized or very small telescopes to search either fainter or
brighter stars.
Five transiting planets 
(OGLE-TR-10b, 56b, 111b, 113b, and 132b) 
have been detected
by the Optical Gravitational Lensing Experiment (OGLE) 
survey ({\em Udalski et al.}, 2002a, 2002b, 2002c, 2003, 2004), 
which uses a 1.3~m telescope.
The parent stars of these planets are faint (typically
$V = 16.5$).
The large-telescope follow-up
observations needed to verify their planetary status, to measure
the stellar reflex velocities, and to estimate the planetary masses
and radii have been conducted by several groups
({\em Bouchy et al.}, 2004, 2005b; {\em Dreizler et al.}, 2002;
{\em Konacki et al.}, 2003a, 2003b, 2004, 2005; {\em Moutou et al.}, 2004;
{\em Pont et al.}, 2004; and {\em Torres et al.}, 2004a, 2004b, 2005).

The Trans-Atlantic Exoplanet Survey (TrES) employed a network
of 3 automated small-aperture (10~cm), wide-field ($6\degr \times 6\degr$) 
telescopes ({\em Brown and Charbonneau}, 2000; {\em Dunham et al.}, 2004; 
{\em Alonso}, 2005) to detect the planet TrES-1 ({\em Alonso et al.}, 2004;
{\em Sozzetti et al.}, 2004).  Its parent star ($V = 11.8$) is significantly
brighter that the OGLE systems, but fainter than the transiting-planet
systems detected by radial-velocity surveys (below). 
Because of this relative accessibility, TrES-1 has also been the 
subject of intensive follow-up observations, as detailed later in this review.

Numerous other photometric transit surveys are active at the current time.
The BEST ({\em Rauer et al.}, 2005), HAT ({\em Bakos et al.}, 2004), 
KELT ({\em Pepper et al.}, 2004), SuperWASP ({\em Christian et al.}, 2005), 
Vulcan ({\em Borucki et al.}, 2001), and XO ({\em McCullough et al.}, 2005) surveys, 
and the proposed PASS ({\em Deeg et al.}, 2004)
survey all adopt the small-aperture, wide-field approach, whereas the 
EXPLORE ({\em Mallen-Ornelas et al.}, 2003) project employs larger telescopes 
to examine fainter stars.  The benefits of surveying stellar
clusters ({\em Janes}, 1996; {\em Pepper and Gaudi}, 2005) have motivated
several surveys of such systems, including EXPLORE/OC ({\em von Braun et al.}, 2005), 
PISCES ({\em Mochejska et al.}, 2005, 2006), and STEPSS ({\em Burke et al.}, 2004; 
{\em Marshall et al.}, 2005).  An early, stunning null result was the
{\em HST} survey of 34,000 stars in the globular cluster 47~Tuc,
which points to the interdependence of the formation and migration of
hot Jupiters on the local conditions, namely
crowding, metallicity, and initial proximity to O and B stars ({\em Gilliland et al.}, 2000).

Finally, three transiting planets were first discovered by 
radial-velocity surveys.
These include HD~209458b, the first transiting planet discovered
({\em Charbonneau et al.}, 2000; {\em Henry et al.}, 2000; {\em Mazeh
et al.}, 2001), and the two most recently discovered
transiting planets, HD~149026b ({\em Sato et al.}, 2005) and HD~189733b
({\em Bouchy et al.}, 2005a).  The latter two objects were uncovered by
quick-look radial-velocity surveys targeted at identifying short-period
planets of metal-rich stars (respectively, the N2K Survey, {\em Fischer et al.}, 2005; 
and the Elodie Metallicity-Biased Search, {\em da Silva et al.}, 2006).
Given the preference of radial-velocity surveys for bright stars,
it is not surprising that all three systems are bright 
($7.6 < V < 8.2$), making them natural targets for detailed follow-up
observations.  As we shall see below, HD~209458b has been extensively studied 
in this fashion.  Similar attention has not yet been lavished on the other two, but only
because of their very recent discovery.
\\

{\em 2.1.3.\ Biases and False Alarms.} Photometric transit surveys 
increase their odds of success by simultaneously 
observing as many stars as possible. 
Hence, their target starfields are moderately to extremely crowded, and the
surveys must therefore work near the boundary of technical
feasibility. The constraints imposed by the search method
influence which kinds of planets are detected.

Photometric transit searches are strongly biased
in favor of planets in small orbits, since such objects
have a greater probability of presenting an eclipsing
configuration (\S2.1.1).  Moreover, most transit searches require
a minimum of 2 (and usually 3) distinct eclipses to be observed, both 
to confirm the reality of the signal, and to permit an evaluation of the orbital period.
Since larger orbits imply longer orbital periods and fewer chances for
transits to occur, small orbits are preferred for transit surveys
with only a limited baseline.  This is frequently the regime in which
single-site surveys operate.  However, multi-site surveys that monitor
a given field for several months (e.g.\ HAT, TrES) frequently achieve
a visibility (the fraction of systems of a given period
for which the desired number of eclipse events would be observed) nearing
100\% for periods up to 6~days.  As a result, such surveys do not suffer
this particular bias, although admittedly only over a limited range of periods.
Similarly, a stroboscopic effect can afflict single-site surveys, favoring 
orbital periods near integer numbers of days
and may account for the tendency of the longer-period
transiting planet periods to clump near 3 and 3.5 days ({\em Pont et al.}, 2004, {\em Gaudi et al.}, 2005).
This situation occurs if the campaign is significantly shorter in duration than that required 
to achieve complete visibility across the desired range or orbital periods.
However, for observing campaigns for which more than adequate phase
coverage has been obtained, the opposite is true, and periods near
integer and half-integer values are disfavored.  The limiting example
of this situation would be a single-site campaign consisting of thousands
of hours of observations, which nonetheless would be insensitive
to systems with integer periods, if their eclipses always occur
when the field is below the horizon.

Most field surveys operate in a regime limited by the
signal-to-noise of their time series (which are typically searched
by an algorithm than looks for statistically-significant, transit-like
events, e.g.\ {\em Kov{\'a}cs et al.}, 2002), and for which the number of
stars increases with decreasing flux (a volume effect).  An important detection bias for surveys
operating under such conditions has been discussed by {\em Pepper et al.\ }(2003) and described in
detail by {\em Gaudi et al.\ }(2005), {\em Gaudi} (2005), and {\em Pont et al.\ }(2005).  
These surveys can more readily detect 
planets with shorter periods and larger radii orbiting fainter stars, and since
such stars correspond to a large distance (hence volume) they are
much more numerous.  As a result, any such survey will reflect this
bias, which cannot be corrected merely by improving the cadence, 
baseline, or precision of the time series (although improving the
latter will reduce the threshold of the smallest planets
that may be detected).

Most ongoing transit surveys are plagued by a high rate of
candidate systems displaying light curves that precisely mimic the desired 
signal, yet are not due to planetary transits.  We can divide
such false positives into three broad categories:  Some are
true {\em statistical} false positives, resulting from selecting
an overly-permissive detection threshold whereby the light-curve
search algorithm flags events that result purely from photometric
noise outliers ({\em Jenkins et al.}, 2002).  The second source is {\em instrumental}, 
due to erroneous photometry, often resulting from leakage 
of signal between the photometric apertures of nearby stars in a crowded field.
However, the dominant form, which we shall term {\em astrophysical} false positives,
result from eclipses among members of double- or multiple-star systems.
Grazing eclipses in binary systems can result in
transit-like signals with depths and durations that resemble planetary
ones ({\em Brown}, 2003), and this effect is especially pronounced
for candidate transits having depths greater than 1\%.
(For equal-sized components, roughly 20\% of eclipsing systems have
eclipse depths that are less than 2\% of the total light.)
In these cases the eclipse shapes are dissimilar (grazing eclipses produce
V-shapes, while planetary transits have flat bottoms), but in
noisy data, this difference can be difficult to detect.
A false alarm may also occur when a small star transits a large one
(e.g., an M-dwarf eclipsing a main-sequence F star).  Since the
lowest-mass stars have Jupiter radii, it is not surprising that
such systems mimic the desired signal closely:
They produce flat-bottomed transits with the correct depths and durations.
Larger stars eclipsing even larger primaries can also
mimic the desired signal, but a careful analysis of the
transit shape can often reveal the true nature of the system
({\em Seager and Mallen-Ornelas}, 2003).
Other useful diagnostics emerge from careful analysis of the light curve
outside of eclipses.
These can reveal weak secondary eclipses, periodic variations due to
tidal distortion or gravity darkening of the brighter component,
or significant color effects.
Any of these variations provides evidence that the eclipsing object has
a stellar mass as opposed to a planetary mass ({\em Drake}, 2003; 
{\em Sirko and Paczy\'nski}, 2003; {\em Tingley}, 2004).
In the absence of these diagnostics, the stellar nature of most
companions is easily revealed by low-precision (1~${\rm km\, s^{-1}}$) radial velocity
measurements, since even the lowest-mass stellar companions cause reflex
orbital motions of tens of ${\rm km\, s^{-1}}$ (for examples, 
see {\em Latham}, 2003; {\em Charbonneau et al.}, 2004;
{\em Bouchy et al.}, 2005b; {\em Pont et al.}, 2005).

The most troublesome systems are hierarchical triple stars in which
the brightest star produces the bulk of the system's light, and the two
fainter ones form an eclipsing binary.
In such cases, the depths of the eclipses are diluted by light from the
brightest member, and often radial velocity observations detect only the bright
component as well.
Given neither radial velocity nor photometric evidence for a binary star,
such cases can easily be mistaken for transiting planets.
Correct identification then hinges on more subtle characteristics of
the spectrum or light curve, such as line profile shapes that vary with
the orbital period ({\em Mandushev et al.}, 2005; {\em Torres et al.}, 2004b, 2005), or color dependence
of the eclipse depth ({\em O'Donovan et al.}, 2006).
Because of the large preponderance of false alarms over true planets,
it is only after all of the above tests have been passed that it makes sense
to carry out the resource-intensive high-precision radial-velocity observations
that establish beyond question that the transiting object has a planetary
mass.
\\

\begin{figure*}
\begin{center}
\includegraphics[angle=90,width=4.5in]{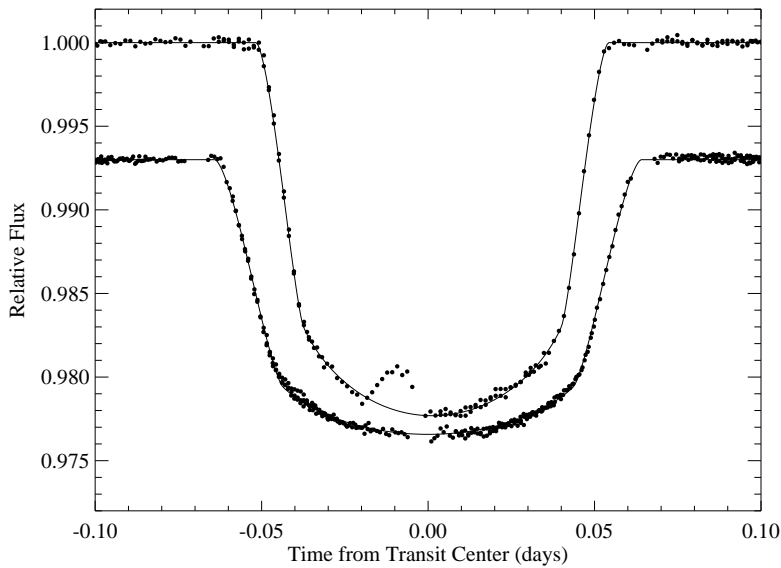}
\caption{\small {\em HST} photometric light curves of transits of TrES-1 (top; {\em Brown et al., in preparation})
and HD~209458 (bottom; {\em Brown et al.} 2001, offset by $-0.007$ for clarity).
The shorter orbital period and the smaller size of the TrES-1 star result in a 
transit that is shorter in duration than that of HD~209458.  Similarly, the smaller 
star creates a deeper transit for TrES-1, despite the fact that HD~209458b is the 
larger planet; the planetary sizes also affect the duration of ingress and egress.  
The TrES-1 data reveal a ``hump'' centered at a
time of $-0.01$~d.  This is likely the result of the planet occulting a starspot 
(or complex of starspots) on the stellar surface.}
\end{center}
\end{figure*}

{\em 2.1.4.\ Determining the Radii and Masses.} After transiting 
planets are identified, an arsenal of observing
tools is available (and necessary) for their characterization.
An accurate estimate of $M_{pl}$ requires 
precise radial-velocity measurements (from which the orbital elements
$P,\ e$, and $\varpi$ are also determined), as well as an estimate of $M_{\star}$. 
The former are gathered 
with high-dispersion echelle spectrographs fed by large telescopes.
For bright parent stars, precision of a few ${\rm m\, s^{-1}}$ (compared to reflex orbital
speeds of $50 - 200\ {\rm m\, s^{-1}}$) can be obtained with convenient exposure times,
so that uncertainties in the velocity measurements do not dominate
the estimate of $M_{pl}$.  In this regime, the greatest source of 
uncertainty is the value of $M_{\star}$ itself.
Given the difficulty of estimating the ages of field stars,
comparison with grids of stellar models (e.g.\ {\em Girardi et al.}, 2002)
suggests that mass estimates are likely to be in error by as much as
5\%.  This uncertainty could be removed by measuring the orbital speed of the
planet directly.  Several efforts have sought to recover the
reflected-light spectrum of the planet in a series of high-resolution
stellar spectra spanning key phases of the orbital period,
but have achieved only upper limits ({\em Charbonneau et al.}, 1999;
{\em Collier Cameron et al.}, 2002; {\em Leigh et al.}, 2003a, 2003b).
(These results also serve to constrain the wavelength-dependent
planetary albedo, a topic to which we shall return in \S 3.2.2.)
For faint parent stars, the radial-velocity estimates become more expensive and problematic, 
and contribute significantly to the final error budget for $M_{pl}$.  
Interestingly, the most intractable uncertainty concerning 
masses of non-transiting planets, namely the value of $\sin i$, is exquisitely
well-determined by fits to the transit light curve.

Analysis of moderate-precision light curves (obtained with ground-based
telescopes) nonetheless yield a tight constraint on the ratio $R_{pl} / R_{\star}$.
However, fits to such data exhibit a fundamental degeneracy
amongst the parameters $R_{pl}$, $R_{\star}$, and $i$, whereby the
planet and stellar radii may be reduced in proportion 
so as to preserve the transit depth, and the orbital inclination may
be correspondingly increased so as to conserve the chord length across the star. 
The uncertainty in $R_{pl}$ is typically dominated by such
degeneracies.  Determining the value of $R_{pl}$ requires fitting eclipse curves
(facilitated by the analytic formulae of {\em Mandel and Agol}, 2002)
subject to independent estimates of $M_{\star}$, $R_{\star}$, 
and the stellar limb-darkening coefficients.
If sufficient photometric precision can be achieved, the value
of $R_{\star}$ may be derived from the light curve itself.
This results in a reduced uncertainty on the value of
$R_{pl}$, due to its weaker dependence on $M_{\star}$,
($\Delta R_{pl} / R_{pl}) \simeq 0.3 (\Delta M_{\star} / M_{\star})$; 
see {\em Charbonneau} (2003).  For illustrative examples 
of the degeneracies that result from such fits,
see {\em Winn et al.\ }(2005), {\em Holman et al.\ }(2005), 
and {\em Charbonneau et al.\ }(2006).

{\em HST} has yielded spectacular transit light curves for two bright systems, 
HD~209458 ({\em Brown et al.}, 2001) and \mbox{TrES-1} ({\em Brown et al., in preparation}), 
which are shown in Fig.~2.
The typical precision of these lightcurves is $10^{-4}$ per one-minute integration,
sufficient to extract new information from relatively subtle
properties of the light curve, such as the duration of the ingress and
egress phases, and the curvature of the light curve near the transit center.
In practice, such data have permitted a simultaneous fit that yields estimates of 
$R_{pl}$, $R_{\star}$, $i$, and the stellar limb-darkening coefficients,
thus reducing the number of assumed parameters to one: $M_{\star}$.
{\em Cody and Sasselov} (2002) point
out that the combined constraint on ($M_{\star}, R_{\star}$) is nearly
orthogonal to that resulting from light-curve fitting, serving
to reduce the uncertainty in $R_{pl}$.  Further improvements
can result from the simultaneously fitting of multi-color photometry under 
assumed values for the stellar-limb darkening, which serves to isolate the
impact parameter (hence $i$) of the planet's path across the star and break the
shared degeneracy amongst $R_{pl}$, $R_{\star}$, and $i$ ({\em Jha et al.}, 2000;
{\em Deeg et al.}, 2001).  Recently, {\em Knutson et al.\ }(2006) 
have analyzed a spectrophotometric {\em HST} dataset spanning $290-1060$~nm, 
and the combined effect of the constraints described above has been to permit the most precise 
determination of an exoplanet radius to date
(HD~209458b; $R_{pl} = 1.320 \pm 0.025~R_{\rm Jup}$).
\\

{\em 2.1.5.\ Further Characterization Measurements.} High-resolution stellar 
spectra obtained during transits can be used to determine
the degree of alignment of the planet's orbital angular momentum vector
with the stellar spin axis. As the planet passes in front of the star,
it produces a characteristic time-dependent shift of the photospheric
line profiles that stems from occultation of part of the rotating
stellar surface. This phenomenon is known as the Rossiter-McLaughlin
effect ({\em Rossiter}, 1924; {\em McLaughlin}, 1924), and has long been 
observed in the spectra of eclipsing binary stars. 
{\em Queloz et al.\ }(2000) and {\em Bundy and Marcy} (2000) detected this effect 
during transits of HD 209458.  A full analytic treatment 
of the phenomenon in the context of transiting
extrasolar planets has been given by {\em Ohta et al.\ }(2005).  
{\em Winn et al.\ }(2005) analyzed the extensive radial-velocity dataset
of HD~209458, including 19 measurements taken during transit.
They found that the measurements of the radial velocity of HD~209458 during
eclipse exhibit an effective half-amplitude of $\Delta v \simeq 55\ {\rm m \, s^{-1}}$, 
indicating a line-of-sight rotation speed of the star of 
$v \sin i_{\star}=4.70 \pm 0.16\ {\rm km \, s^{-1}}$. They also detected
a small asymmetry in the Rossiter-McLaughlin anomaly, which they
modeled as arising from an inclination, $\lambda$, of the planetary orbit relative to the
apparent stellar equator of $\lambda=-4.4^{\circ} \pm 1.4^{\circ}$. Interestingly,
this value is smaller than the $\lambda=7^{\circ}$ tilt of the solar rotation axis
relative to the net angular momentum vector defined by the 
orbits of the solar system planets (see {\em Beck and Giles}, 2005). 
{\em Wolf et al.\ }(2006) carried out a similar analysis for HD~149026, and found
$\lambda=12^{\circ} \pm 14^{\circ}$. For these planets, the timescales for tidal
coplanarization of the planetary orbits and stellar equators are
expected to be of order $10^{12}\ {\rm yr}$ ({\em Winn et al.}, 2005;
{\em Greenberg}, 1974; {\em Hut}, 1980), indicating that the observed value of 
$\lambda$ likely reflects that at the end of the planet formation process.

Pertubations in the timing of planetary transits may be used to infer the
presence of satellites or additional planetary companions ({\em Brown et al.}, 2001;
{\em Miralda-Escud{\'e}}, 2002). {\em Agol et al.\ }(2005) and {\em Holman and Murray} (2005) 
have shown how non-transiting terrestrial-mass planets could be detected
through timing anomalies.  Although {\em HST} observations have yielded the most
precise timing measurements to date (with a typical precision of 10s; 
see tabulation for HD~209458 in {\em Wittenmyer et al.}, 2005), 
the constraints from ground-based observations can nonetheless 
be used to place interesting limits on additional planets in the system,
as was recently done for TrES-1 ({\em Steffen and Agol}, 2005).

Precise photometry can also yield surprises, as in the 
``hump'' seen in Fig.~2.  This
feature likely results from the planet crossing a large sunspot
(or a complex of smaller ones), and thus is evidence for magnetic
activity on the surface of the star.
Such activity may prove to be an important noise source for
timing measurements of the sort just described,
but it is also an interesting object of study in its own right,
allowing periodic monitoring of the stellar activity
along an isolated strip of stellar latitude ({\em Silva}, 2003).

\subsection{\textbf{Theory and Interpretation}}

\begin{deluxetable}{llllllllll}
\tabletypesize{\small}
\tablecaption{Properties of the Transiting Planets \label{tbl-1}}
\tablewidth{0pt}
\tablehead{Planet & $M_{\star}$ & $P$ & $T_{{\rm eff},\star}$ &
$R_{\star}$ & $R_{pl}$ & $M_{pl}$
& $T_{{\rm eq},pl}$ & $R_{pl}$ & $R_{pl}$\\
 & $M_{\odot}$ & days & K & $R_{\odot}$ & $R_{\rm
Jup}$ & $M_{\rm Jup}$ & K & $20 M_{\oplus}\,$core & no core}
\startdata
OGLE-TR-56b  & 1.04$\pm$.05 & 1.21 & 5970$\pm$150 & 1.10$\pm$.10 & 1.23$\pm$.16 & 1.45$\pm$.23 & 1800$\pm$130 & 1.12$\pm$.02 & 1.17$\pm$.02\\
OGLE-TR-113b & 0.77$\pm$.06 & 1.43 & 4752$\pm$130 & 0.76$\pm$.03 & $1.08^{+.07}_{-.05}$ & 1.35$\pm$.22 & 1186$\pm$\phantom{1}78 & 1.07$\pm$.01 & 1.12$\pm$.01\\
OGLE-TR-132b & 1.35$\pm$.06 & 1.69 & 6411$\pm$179 & 1.43$\pm$.10 & 1.13$\pm$.08 & 1.19$\pm$.13 & 1870$\pm$170 & 1.13$\pm$.02 & 1.18$\pm$.02\\
HD~189733b & 0.82$\pm$.03 & 2.22 & 5050$\pm$\phantom{0}50 & 0.76$\pm$.01 & 1.15$\pm$.03 & 1.15$\pm$.04 & 1074$\pm$\phantom{1}58 & 1.07$\pm$.01 & 1.11$\pm$.01\\
HD~149026b & 1.30$\pm$.10 & 2.88 & 6147$\pm$\phantom{0}50 & 1.45$\pm$.10 & 0.73$\pm$.05 & 0.36$\pm$.03 & 1533$\pm$\phantom{1}99 & 0.98$\pm$.02 & 1.15$\pm$.02\\
TrES-1    & 0.87$\pm$.03 & 3.00 & 5214$\pm$\phantom{0}23 & 0.83$\pm$.03 & 1.12$\pm$.04 & 0.73$\pm$.04 & 1038$\pm$\phantom{1}61 & 1.02$\pm$.01 & 1.10$\pm$.00\\
OGLE-TR-10b  & 1.00$\pm$.05 & 3.10 & 6220$\pm$140 & 1.18$\pm$.04 & 1.16$\pm$.05 & 0.54$\pm$.14 & 1427$\pm$\phantom{1}88 & 1.01$\pm$.02 & 1.13$\pm$.01\\
HD~209458b & 1.06$\pm$.13 & 3.52 & 6099$\pm$\phantom{0}23 & 1.15$\pm$.05 & 1.32$\pm$.03 & 0.66$\pm$.06 & 1314$\pm$\phantom{1}74 & 1.02$\pm$.01 & 1.12$\pm$.00\\
OGLE-TR-111b & $0.82^{+.15}_{-.02}$ & 4.02 & 5070$\pm$400 & $0.85^{+.10}_{-.03}$ & $1.00^{+.13}_{-.06}$ & 0.53$\pm$.11 & \phantom{1}930$\pm$100 & 0.97$\pm$.02 & 1.09$\pm$.01\\
\enddata
\end{deluxetable}

{\em 2.2.1.\ Overview and Uncertainties.} Transiting planets give us the opportunity to test our
understanding of the physical structure of giant planets.  In particular,
structural models of the known
transiting planets must be able to account for the wide range of radiation fluxes to
which these planets are subjected, and they must recover the observed
range of radii.
In general, as the planetary mass decreases,
a given external energy input has an increasingly larger influence on the 
size and interior structure of the planet.
For hot Jupiters, the absorbed stellar flux
creates a radiative zone in the subsurface regions that 
controls the planetary contraction, and ultimately dictates the radius.
Models of transiting giant planets straddle the physical characteristics of brown dwarfs
and low-mass stars, as
well as the solar system giants 
(for an overall review, see {\em Burrows et al.}, 2001).

The construction of structural models for giant planets
is difficult because a number of key physical inputs are
poorly constrained. This situation holds equally for extrasolar planets 
and for the exquisitely observed outer planets of the Solar system.
A benefit of robust determinations of the parameters
for a growing range
of planets is that uncertain aspects of the theory 
can become increasingly constrained. Indeed,
transit observations have the potential to clarify some of
the core questions regarding giant planets.

The dominant uncertainty regarding the overall structure of gas giants
is in the equation of state (see the review of
{\em Guillot}, 2005). The interiors of solar system and extrasolar
giant planets consist of partially degenerate, partially ionized
atomic-molecular fluids ({\em Hubbard}, 1968). The pressure, $P$,
in the interiors of most giant planets exceeds 10~Mbar, and central
temperatures range from $T_{c} \simeq 10^{4}$ for Uranus and Neptune
to $T_{c} \simeq 3\times10^{4}$ for objects such as HD~209458b. This material regime
lies beyond the point where hydrogen ionizes and becomes metallic, although
the details of the phase transition are still uncertain ({\em Saumon et al.}, 2000;
{\em Saumon and Guillot}, 2004). The equation of state of giant planet interiors
is partially accessible to laboratory experiments, including gas-gun
({\em Holmes et al.}, 1995),
laser-induced shock compression ({\em Collins et al.}, 1998), pulsed-power shock compression
({\em Knudson et al.}, 2004), and convergent shock wave ({\em Boriskov et al.}, 2003)
techniques. These experiments can achieve momentary pressures in excess of 1~Mbar,
and they appear to be approaching the molecular to metallic hydrogen transition.
Unfortunately, these experiments report diverging results.  In particular, they yield a range
of hydrogen compression factors relevant to planetary cores that differ by $\sim50\%$.
Furthermore, the laboratory experiments are in only partial agreement with
first-principles quantum mechanical calculations of the hydrogen equation of 
state ({\em Militzer and Ceperley}, 2001; {\em Desjarlais}, 2003; {\em Bonev et al.}, 2004),
and uncertainties associated with the equations of state of helium and heavier elements
are even more severe ({\em Guillot}, 2005).
At present, therefore, structural models must adopt the 
pragmatic option of choosing a thermodynamically
consistent equation of state that reproduces either the
high- or low-compression results ({\em Saumon and Guillot}, 2004).

Another uncertainty affecting the interior models is the existence 
and size of a radial region where helium separates from hydrogen and forms
downward-raining droplets. The possibility that giant planet interiors are helium-stratified
has non-trivial consequences for their structures, and ultimately, their sizes.
In the case of Saturn, the zone of helium 
rain-out may extend all the way to the center, possibly resulting in a distinct
helium shell lying on top of a heavier element core ({\em Fortney and Hubbard}, 2003).

Uncertainties in the equation of state, the bulk composition, and the degree of
inhomogeneity allow for a depressingly wide range of models for the solar system giants that 
are consistent with the observed radii, surface temperatures, and gravitational moments. In
particular ({\em Saumon and Guillot}, 2004), one can construct observationally-consistent
models for Jupiter with core masses ranging from $0 - 12~M_{\oplus}$, 
and an overall envelope heavy-element content ranging from $6 - 37~M_{\oplus}$. 
This degeneracy must be broken in order to
distinguish between the core accretion ({\em Mizuno}, 1980; {\em Pollack et al.}, 1996; 
{\em Hubickyj et al.}, 2004) and gravitational instability ({\em Boss}, 1997, 2000, 2004)
hypotheses for planet formation.
Fortunately, the growing dataset of observed masses and radii from the
transiting extrasolar planets suggests a possible strategy for resolving the
tangle of uncertainties.
The extreme range of temperature conditions under which hot Jupiters exist, along
with the variety of masses that are probed, can potentially provide definitive
constraints on the interior structure of these objects.
\\

{\em 2.2.2.\ Comparison to Observations.} Following the discovery
of 51~Pegb ({\em Mayor and Queloz}, 1995), models of Jovian-mass planets
subject to strong irradiation were computed ({\em Lin, Bodenheimer and
Richardson}, 1996; {\em Guillot et al.}, 1996). These models predicted that
short-period Jovian-mass planets with effective temperatures of roughly 
$1200$~K would be significantly larger than Jupiter, and the
discovery that HD~209458b has a large radius initially
seemed to confirm these calculations. In general, $R_{pl}$ is a
weak function of planet mass, reflecting the overall $n=1$
polytropic character of giant planets ({\em Burrows et al.}, 1997, 2001).

In order to evaluate the present situation, we have collected 
the relevant quantities for the 9 transiting planets in Table~1. 
In particular, we list the most up-to-date estimates of 
$P$, $R_{\star}$, $M_{\star}$, $R_{pl}$, $M_{pl}$,
as well as the stellar effective temperature, $T_{{\rm eff}, \star}$. 
We also list the value of the planetary equilibrium temperature, $T_{{\rm eq}, pl}$,
which is calculated
by assuming the value for the Bond albedo, $A$, recently estimated
for TrES-1 ($A=0.31 \pm 0.14$; {\em Charbonneau et al.}, 2005; \S3.2.3).
The precision of the estimates of the physical properties varies considerably
from star to star.  By drawing from the Gaussian distributions corresponding to the 
uncertainties in Table~1 and the quoted value for $A$, we can estimate the
uncertainty for $M_{pl}$ and $T_{{\rm eq}, pl}$ for each planet. Thereafter, for a 
particular choice of $M_{pl}$ and $T_{{\rm eq}, pl}$, and fixing the planetary age at 4.5~Gyr, 
we can compute theoretical radii.
For this task, we use the results
of {\em Bodenheimer et al.\ }(2003), who computed models for insolated
planets ranging in mass from $0.11-3.0\ M_{\rm Jup}$. To
evaluate the radii differences that arise from different heavy element fractions, separate
sequences were computed for models that contain and do not contain 
20-$M_{\oplus}$ solid cores, and both predictions are
listed in Table~1.  The models have been calibrated so that, for the evolution of Jupiter up to the age of
4.5~Gyr, a model with a core gives the correct Jupiter radius to within 1\%.
Planetary age can also have a significant effect on $R_{pl}$. 
For example, the evolutionary models of {\em Burrows et al.\ }(2004) for 
OGLE-TR-56b ($1.45\ M_{\rm Jup}$) yield transit radii of
$R_{pl} \simeq 1.5\ R_{\rm Jup}$ at 100~Myr, and $\sim1.25\ R_{\rm Jup}$ 
after 2~Gyr. In general, however, $R_{pl}$ evolves only modestly beyond the
first 500~Myr, and hence the uncertainties in the ages of the
parent stars (for which such young ages may generally be excluded)
introduce errors of only a few percent into the values of $R_{pl}$.

The models use a standard Rosseland mean photospheric boundary condition, and
as such, are primarily intended for cross-comparison of radii.
The obtained planetary radii are, however,
in excellent agreement with baseline models obtained by groups
employing detailed frequency-dependent atmospheres (e.g.\ {\em Burrows et al.}, 2004;
{\em Chabrier et al.}, 2004; {\em Fortney et al.}, 2005b). 
The models assume that the
surface temperature is uniform all the way around the planet, even though
the rotation of the planet is likely tidally locked.
Hydrodynamic simulations of the atmosphere that aim, in part, to evaluate
the efficiency with which the planet redistributes heat from the dayside to the nightside 
have been performed by {\em Cho et al.\ }(2003), {\em Showman and Guillot} (2002),
{\em Cooper and Showman} (2005), and {\em Burkert et al.\ }(2005) under various simplifying
assumptions. There is no agreement on what the temperature difference between the 
dayside and the nightside should be (\S3.2.4), and it 
depends on the assumed opacity in the atmosphere. {\em Burkert et al.\ }(2005) 
suggest that with a reasonable opacity, the difference could be $200$~K,
not enough to make an appreciable difference in the radius.

A number of interesting conclusions regarding the bulk
structural properties of the transiting planets can be drawn from Table~1. 
First, the baseline radius predictions display ($1\sigma$) agreement 
for seven of the nine known transiting planets. 
Second, the planets whose radii are in good agreement
with the models span the full range of masses and effective temperatures. The models do not 
appear to be systematically wrong in some particular portion of parameter
space.  Although the reported accuracies of the basic 
physical parameters are noticeably worse
for the OGLE systems than for the brighter targets, the constraints 
are nonetheless useful to address models of their physical structure and,
in particular, the presence or absence of a solid core.
Specifically, the baseline models in Table~1 indicate that the presence
of a solid core in a $0.5 \, M_{\rm Jup}$ planet with \tp$\, =1500$~K leads to a 
radius reduction of roughly $0.1 \, R_{\rm Jup}$. This difference
generally exceeds the uncertainty in the estimate of $R_{pl}$.

In the standard core-accretion paradigm for giant planet
formation, as reviewed by {\em Lissauer} (1993),
a Jovian planet arises from the
collisional agglomeration of a solid 10-$M_{\oplus}$ core over a period
of several million years, followed by a rapid accretion of hundreds of Earth masses 
of nebular gas, lasting roughly $10^5$~yr. The competing
gravitational instability hypothesis (e.g.\ {\em Boss} 1997, 2004) posits that
gas-giant planets condense directly from spiral instabilities in protostellar
disks on a dynamical timescale of less than $10^3$~yr.
{\em Boss} (1998) points out that solid particles in the newly formed planet can
precipitate to form a core during the initial contraction phase.
Only 1\% of the matter in the planet is condensible, however,
so a Jovian-mass planet that formed by this process will have a core that is
much less massive than one that formed by the core-accretion scenario.

Among the 7 planets that show agreement with the baseline models,
it is presently difficult to discern the presence of a core.
However, the ``transit radius" effect ({\em Burrows et al.}, 2003; \S2.2.3)
will tend to systematically increase the
observed radii above the model radii listed in Table~1 (which correspond to
a 1-bar pressure level). Similarly, 
signal-to-noise-limited field transit surveys bias 
the mean radius of planets so detected to a value larger than that of the intrinsic population
({\em Gaudi}, 2005).  Taking both effects into account lends favor to the models with cores.
Clearly more transiting planets and more precise determinations of their properties
are necessary, as are more physically detailed models.  We note that the identification
of lower-mass transiting planets (for which the effect of a solid
core is prominent) would be particularly helpful to progress in these questions.
Several groups (e.g.\ {\em Gould et al.}, 2003; {\em Hartman et al.}, 2005; {\em Pepper and Gaudi}, 2006)
have considered the prospects for ground-based searches for planets with radii of
that of Neptune, or less.
\\

\begin{figure*}
\begin{center}
\includegraphics[width=6.0in]{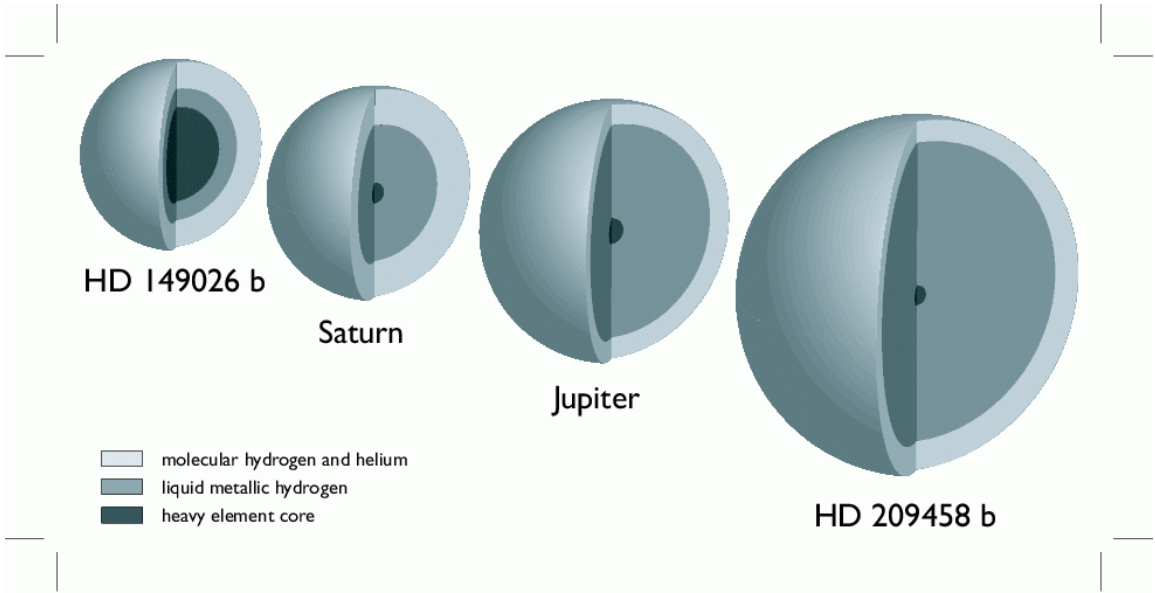}
\end{center}
\caption{\small Cut-away diagrams of Jupiter, Saturn, and the two oddball extrasolar
planets, drawn to scale.  The observed radius of HD~149026b implies a massive core of heavy elements
that makes up perhaps 70\% of the planetary mass.  In contrast, the radius of
HD~209458b intimates a coreless structural model, as well as an additional
energy source to explain its large value.}
\end{figure*}

{\em 2.2.3.\ The Transit Radius Effect.} When a planet occults its parent star, the wavelength-dependent
value of $R_{pl}$ so inferred is not necessarily  the canonical planetary radius at a
pressure level of 1~bar ({\em Lindal et al.}, 1981; {\em Hubbard et al.}, 2001),
which we have used for the baseline predictions in Table~1.  
As such, the measured radius is 
approximately the impact parameter of the transiting planet at which the optical depth to the stellar 
light along a chord parallel to the star-planet line of centers is unity. This
is not the optical depth in the radial direction, nor is it associated with the radius at the
radiative-convective boundary.   Hence, since the pressure level to which the transit beam is probing
near the planet's terminator is close to one {\it milli}bar ({\em Fortney et al.}, 2003),
there are typically $5-10$ pressure scale heights difference between the measured value of $R_{pl}$
and either the radiative-convective boundary ($\ge$1000 bars) and the 1-bar radius.  
(If, as discussed in {\em Barman et al.\ }(2002), the transit radius is at pressures well below the 1~millibar 
level, then the effect would be even larger.)
Furthermore, exterior to the radiative-convective boundary,
the entropy is an increasing function of radius.  One consequence
of this fact is significant radial inflation
vis \`a vis a constant entropy atmosphere.  Both of these effects result in an
apparent increase of perhaps $0.1\  R_{\rm Jup}$ ($\sim$7\%) in the theoretical 
radius for HD~209458b and $0.05\ R_{\rm Jup}$ ($\sim$4\%) for OGLE-TR-56b.
\\

{\em 2.2.4.\ Explaining the Oddballs.} Two of the planets, HD~209458b and HD~149026b, have
radii that do not agree at all with the predictions. HD~209458b 
is considerably larger than predicted, and
HD~149026b is too small. These discrepancies indicate that the physical
structures of the transiting planets can depend significantly on factors 
other than $M_{pl}$ and \tp. It would appear
that hot Jupiters are imbued with individual personalities.

While the radius of HD~209458b is certainly
broadly consistent with a gas-giant composed primarily of hydrogen,
studies by {\em Bodenheimer et al.\ }(2001) and {\em Guillot and Showman} (2002)
were the first to make it clear that a standard model of a
contracting, irradiated planet can recover $R_{pl} \simeq 1.35\, R_{\rm Jup}$
for HD 209458b only if the deep atmosphere is unrealistically hot. 
A number of resolutions to this conundrum have
been suggested. {\em Bodenheimer et al.\ }(2001) argue
that HD~209458b might be receiving interior tidal heating through ongoing
orbital circularization. This hypothesis was refined by {\em Bodenheimer
et al.\ }(2003), who computed grids of predicted planetary sizes
under a variety of conditions, and
showed that the then-current radial velocity data set for HD~209458b
was consistent with the presence of an undetected planet
capable of providing the requisite eccentricity forcing. The tidal-heating
hypothesis predicts that HD~209458b is caught up in an anomalous situation, and 
that the majority of hot Jupiter-type planets will have considerably smaller
radii than that observed for HD~209458b. Recent analyses by
{\em Laughlin et al.\ }(2005b) and {\em Winn et al.\ }(2005) indicate that the orbital
eccentricity of HD~209458b is close to zero.  This conclusion is
further buttressed by the timing of the secondary eclipse by {\em Deming et al.\ }(2005a;
discussed in greater detail in \S3.2.3), which places stringent upper limits on the eccentricity, 
except in the unlikely event that the orbit is precisely aligned to our line of sight.
Thus the eccentricity appears to be below the value required
to generate sufficient tidal heating to explain the inflated radius.

{\em Guillot and Showman} (2002) proposed an alternate hypothesis in which
strong insolation-driven weather patterns
on the planet drive the conversion of kinetic energy into
thermal energy at pressures of tens of bars.
They explored this idea by modifying their planet evolution
code to include a radially adjustable internal energy source term.
They found that if kinetic
wind energy is being deposited at adiabatic depths with an efficiency of
1\%, then the large observed radius of the planet can be explained.
Their hypothesis predicts that other transiting planets with
similar masses and at similar irradiation levels should be
similar in size to HD~209458b. The subsequent discovery that TrES-1 has a
considerably smaller radius despite its similar 
temperature, mass, and parent star metallicity is evidence against the kinetic 
heating hypothesis, since it is not clear why this mechanism should 
act upon only HD~209458b.

Recently, an attractive mechanism for explaining the planet's
large size has been advanced by {\em Winn and Holman} (2005) who suggest that
the anomalous source of heat arises from obliquity tides that
occur as a result of the planet being trapped in a Cassini state (e.g.\ {\em Peale}, 1969). 
In a Cassini state, a planet that is formed with a non-zero
obliquity is driven during the course of spin synchronization to a final state 
in which spin precession resonates with orbital precession. When caught in 
a Cassini state, the planet is forced to maintain a non-zero obliquity,
and thus experiences continued tidal dissipation as a result of orbital
libration. Order-of-magnitude estimates indicate that the amount of expected
tidal dissipation could generate enough heat
to inflate the planet to the observed size.

HD~149026b presents a problem that is essentially the opposite to that
of HD~209458b. Both the mass ($0.36\ M_{\rm Jup}$) and the
radius ($0.73\ R_{\rm Jup}$) are considerably smaller than those of the
other known transiting extrasolar planets.  Curiously, HD~149026 is
the only star of a transiting planet to have a metallicity that
is significantly supersolar, [Fe/H]$\, = 0.36$. The observed radius is 
30\% smaller than the value predicted by the baseline model with a 
core of $20\, M_{\oplus}$. 
Clearly, a substantial enrichment in heavy elements above solar composition
is required. The mean density of the planet, $1.17\ {\rm g\, cm^{-3}}$, is 
1.7$\times$ that of Saturn, which itself has roughly 25\%
heavy elements by mass. On the other hand, the planet is not composed entirely
of water or silicates, or else the radius would be of order 0.4 or 
0.28~$R_{\rm Jup}$, respectively ({\em Guillot et al.}, 1996, {\em Guillot} 2005). 
Models by {\em Sato et al.\ }(2005) and by {\em Fortney et al.\ }(2005b) agree that the 
observed radius can be recovered if the planet contains approximately 
$70\, M_{\oplus}$ of heavy elements, either distributed throughout the interior 
or sequestered in a core.

The presence of a major fraction of heavy elements in HD~149026b has a number
of potentially interesting ramifications for the theory of planet formation.
{\em Sato et al.\ }(2005) argue that it would be difficult to form this giant planet
by the gravitational instability mechanism ({\em Boss}, 2004). The large core also
presents difficulties for conventional models of core accretion. In the core-accretion 
theory, which was developed in the context of the minimum-mass
solar nebula, it is difficult to prevent runaway gas accretion from
occurring onto cores more massive than $30 M_{\oplus}$, even if abundant
in-falling planetesimals heat the envelope and delay the Kelvin-Helmholtz
contraction that is required to let more gas into the planet's Hill sphere.
The current structure of HD~149026b suggests that it was formed in a gas-starved
environment, yet presumably enough gas was present in the protoplanetary disk
to drive migration from its probable formation region beyond $1-2$~AU 
from the star inward to the current orbital separation of 0.043~AU.  Alternately, 
a metal-rich disk would likely be abundant in planetesimals, which may in turn have
promoted the inward migration of the planet via planetesimal scattering
({\em Murray et al.}, 1998).

\section{\textbf{ATMOSPHERES}}

By the standards of the Solar system, the atmospheres of the close-in planets listed 
in Table~1 are quite exotic.  Located only 0.05~AU from their parent stars, these gas giants
receive a stellar flux that is typically $10^{4}$ that which strikes
Jupiter.  As a result, a flurry of theoretical activity over the past
decade has sought to predict (and, more recently, interpret) the emitted and
reflected spectra of these objects (e.g.\ {\em Seager and Sasselov}, 1998; 
{\em Seager et al.}, 2000; {\em Barman et al.}, 2001; {\em Sudarsky et al.}, 2003; {\em Allard et al.}, 2003; 
{\em Burrows et al.}, 2004; {\em Burrows}, 2005).  Observations promise to grant answers to
central questions regarding the atmospheres of the planets, including the identity
of their chemical constituents, the presence (or absence) of clouds, the fraction of
incident radiation that is absorbed (and hence the energy budget of the
atmosphere), and the ability of winds and weather patterns to redistribute
heat from the dayside to the nightside.    For a detailed review of the theory
of extrasolar planet atmospheres, see the chapter by {\em Marley et al}.  We
summarize the salient issues below (\S3.1), and then proceed to discuss the
successful observational techniques, and resulting constraints to date (\S3.2).

\subsection{Theory}

{\em 3.1.1.\ Overview.} In order to model the atmospheres and spectra of extrasolar 
giant planets in general, and hot Jupiters in particular, one must assemble 
extensive databases of molecular and atomic opacities. The species of most 
relevance, and which provide diagnostic signatures, are H$_2$O, CO, CH$_4$, 
H$_2$, Na, K, Fe, NH$_3$, N$_2$, and silicates.  The chemical 
abundances of these and minority species are derived using thermochemical
data and minimizing the global free energy.  Non-equilibrium effects
in the upper atmospheres require chemical networks and kinetic
coefficients.  With the abundances and opacities, as well as 
models for the stellar spectrum, one can embark upon calculations
of the atmospheric temperature, pressure, and composition profiles and 
of the emergent spectrum of an irradiated planet.  
With atmospheric temperatures in the
$1000-2000$~K range, CO, not CH$_4$, takes up much of the carbon in the low-pressure outer
atmosphere, and N$_2$, not NH$_3$, sequesters most of the nitrogen.
However, H$_2$O predominates in the atmospheres for both hot and cooler giants.
Perhaps most striking in the spectrum of a close-in giant planet
is the strong absorption due to the sodium and potassium resonance doublets.  These lines are
strongly pressure-broadened and likely dominate the visible
spectral region. The major infrared spectral features are due to H$_2$O, CO, CH$_4$, 
and NH$_3$.  H$_2$ collision-induced absorption contributes very broad features in the infrared.

A self-consistent, physically realistic evolutionary calculation
of the radius, \tp, and spectrum of a giant planet in isolation requires an outer boundary condition
that connects radiative losses, gravity ($g$), and core entropy ($S$).  
When there is no irradiation, the effective temperature determines both the flux from 
the core and the entire object.
A grid of \tp, $g$, and $S$, derived from detailed atmosphere calculations,
can then be used to evolve the planet (e.g.\ {\em Burrows et al.}, 1997; {\em Allard et al.}, 1997).
However, when a giant planet is being irradiated by its star, this procedure must
be modified to include the outer stellar flux in the calculation that yields
the corresponding $S$-\tp{-$g$} relationship.  This must be done for a given external stellar
flux and spectrum, which in turn depends upon the stellar luminosity spectrum and the 
orbital distance of the giant planet.   Therefore, one needs to calculate a new $S$-\tp{-$g$} 
grid under the irradiation regime of the hot Jupiter that is tailor-made for the luminosity 
and spectrum of its primary and orbital distance.
With such a grid, the radius evolution of a hot Jupiter can be calculated, with
its spectrum as a by-product.
\\

{\em 3.1.2.\ The Day-Night Effect and Weather.} A major issue is
the day-night cooling difference.  The 
gravity and interior entropy are the same for the day and the
night sides.  For a synchronously rotating hot Jupiter,
the higher core entropies needed to explain a large measured radius imply higher internal fluxes
on a night side if the day and the night atmospheres are not coupled 
(e.g.\ {\em Guillot and Showman}, 2002). 
For strongly irradiated giant planets, there is a pronounced inflection and
flattening in the temperature-pressure profile that is predominantly
a result of the near balance at some depth between countervailing incident and internal fluxes.
The day-side core flux is suppressed by this flattening of
the temperature gradient and the thickening of the radiative zone due to irradiation.
However, {\em Showman and Guillot} (2002), {\em Menou et al.\ }(2003), {\em Cho et al.\ }(2003), 
{\em Burkert et al.\ }(2005), and {\em Cooper and Showman} (2005) have recently demonstrated
that strong atmospheric circulation currents that advect heat from the day to the night
sides at a wide range of pressure levels are expected for close-in giant planets.  
{\em Showman and Guillot} (2002)
estimate that below pressures of 1~bar the night-side cooling of the air
can be quicker than the time it takes the winds to traverse the night side, 
but that at higher pressures
the cooling timescale is far longer.  Importantly, the radiative-convective boundary
in a planet such as HD~209458b is very deep, at pressures of perhaps 1000~bar.  This may mean that
due to the coupling of the day and the night sides via strong winds at depth, the temperature-pressure
profiles at the convective boundary on both sides are similar.
This would imply that the core cooling rate is roughly the same in both hemispheres.
Since the planet brightness inferred during secondary eclipse (\S3.2.3) 
depends upon the advection of stellar heat to the night side, such data can
provide onstraints on the meteorology and general circulation models.

Almost complete redistribution of heat occurs in 
the case of Jupiter, where the interior flux is latitude- and
longitude-independent.
However, the similarity in Jupiter of the day and night temperature-pressure profiles and effective
temperatures is a consequence not of the redistribution of heat by rotation
or zonal winds, but of the penetration into the convective zone
on the day side of the stellar irradiation ({\em Hubbard}, 1977).  Core convection
then redistributes the heat globally and accounts for the uniformity of 
the temperature over the entire surface.
Therefore, whether direct heating of the convective zone by the stellar light is responsible,
as it is in our own Jovian planets, for the day-night smoothing can depend on
the ability of the stellar insolation to penetrate below the radiative-convective boundary.
This does not happen for a hot Jupiter.
Clearly, a full three-dimensional study will be required to definitively
resolve this thorny issue.

Clouds high in the atmospheres of hot Jupiters 
with {\tp}$>1500$~K would result in wavelength-dependent flux variations.
Cloud opacity tends to block the flux windows
between the molecular absorption features, thereby reducing the
flux peaks.  Additionally,
clouds reflect some of the stellar radiation,
increasing the incident flux where the scattering opacity is high.
This phenomenon tends to be more noticeable in the vicinity of the gaseous
absorption troughs.

\subsection{Observations}
The direct study of extrasolar planets
orbiting mature (Gyr) Sun-like stars may proceed without the need to image 
the planet (i.e.\ to spatially separate the light of the planet from that of the star).
Indeed, this technical feat has not yet been accomplished.  Rather, the
eclipsing geometry of transiting systems permits the spectrum of the
planet and star to be disentangled through monitoring of the variation
in the combined system light as a function of the known orbital
phase.  Detections and meaningful upper limits have been achieved
using the following three techniques.
\\

\begin{figure*}
\begin{center}
\includegraphics[angle=0.,width=6.2in]{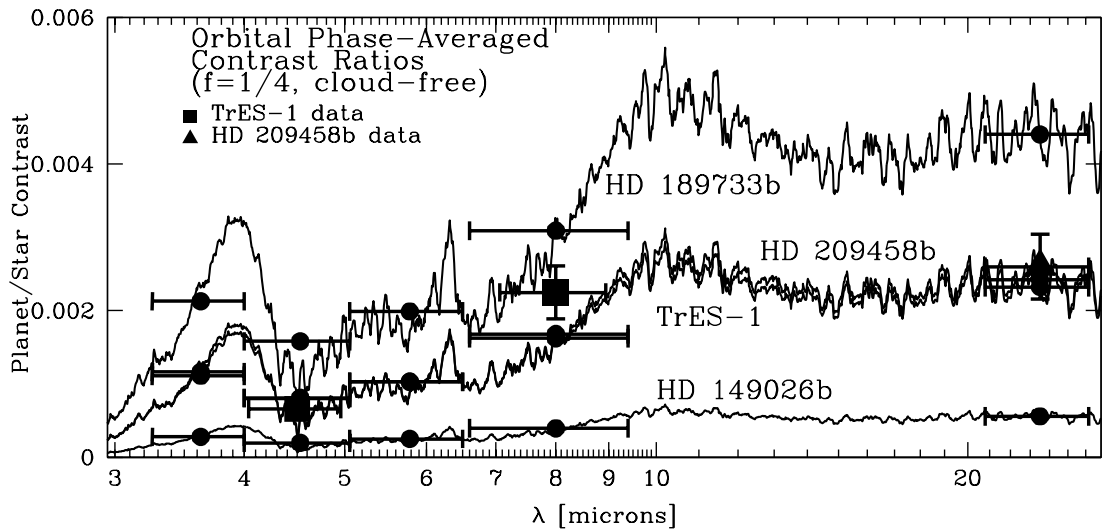}
\caption{\small The orbital phase-averaged planet-to-star flux-density ratio as a function of wavelength 
($\lambda$, in $\mu$m) for the models of the four known transiting extrasolar planets
for which such observations might be feasible ({\em Burrows et al.}, 2005, and {\em in preparation}).
The bandpass-integrated predicted values
are shown as filled circles, with the bandwidths indicated by horizontal
bars.  The measured values for TrES-1 at 4.5~$\mu$m and 8.0~$\mu$m ({\em Charbonneau et al.}, 2005) 
are shown as filled squares, and the observed value at 24~$\mu$m for HD~209458b ({\em Deming et al.}, 2005a)
is shown as a filled triangle.  The extremely favorable contrast for HD~189733,
and the extremely challenging contrast ratio for HD~149026, both result primarily
from the respective planet-to-star surface area ratios.}
\label{fig:4}
\end{center}
\end{figure*}

{\em 3.2.1.\ Transmission Spectroscopy.} The technique of transmission spectroscopy seeks to ratio 
stellar spectra gathered during transit with those taken just before or
after this time, the latter providing a measurement of the spectrum of
the isolated star.  Wavelength-dependent sources of opacity in the upper
portions of the planetary atmosphere, or in its exosphere, 
will impose absorption features that could be revealed in this ratio.
This technique can be viewed as probing the wavelength-dependent variations in the
inferred value of $R_{pl}$.

The first composition signature ({\em Charbonneau et al.}, 2002) was detected with the 
{\em HST} STIS spectrograph.  The team measured an increase in the transit
depth of $(2.32\pm0.57) \times 10^{-4}$ for HD~209458 in a narrow bandpass centered on the
sodium resonance lines near 589~nm.  Ruling out alternate explanations
of this diminution, they conclude that the effect results from absorption due
to atomic sodium in the planetary atmosphere, which indeed had been
unanimously predicted to be a very prominent feature at visible wavelengths 
({\em Seager and Sasselov}, 2000; {\em Hubbard et al.}, 2001; {\em Brown}, 2001).
Interestingly, the detected amplitude was roughly 1/3 that predicted by
baseline models that incorporated a cloudless atmosphere and a solar 
abundance of sodium in atomic form. 
{\em Deming et al.\ }(2005b) follow the earlier work of {\em Brown et al.\ }(2002)
to achieve strong upper limits on the CO bandhead at 2.3~$\mu$m.  Taken together,
the reduced amplitude of the sodium detection and the upper limits on CO
suggest the presence of clouds high in the planetary atmosphere (e.g.\ 
{\em Fortney et al.}, 2003), which
serve to truncate the effective size of the atmosphere viewed in transmission.
{\em Fortney} (2005) considers the slant optical depth and
shows that even a modest abundance of condensates or
hazes can greatly reduce the size of absorption features measured by this
technique.  Alternately, non-LTE effects may explain the weaker-than-expected
sodium feature ({\em Barman et al.}, 2005).

Planetary exospheres are amenable to study by this method, as the increased
cross-sectional area (compared to the atmospheres) implies a large
potential signal.  {\em Vidal-Madjar et al.\ }(2003) observed a $15\pm4$\%
transit depth of HD~209458 when measured at Ly$\alpha$.  The implied
physical radius exceeds the Roche limit, leading them to conclude that
material is escaping the planet ({\em Lecavelier des Etangs et al.}, 2004; {\em Baraffe et al.}, 2004).
However, the minimum escape rate required
by the data is low enough to reduce the planetary mass by only 0.1\%
over the age of the system.  More recently, {\em Vidal-Madjar et al.\ }(2004)
have claimed detection of other elements, with a lower statistical
significance.  Significant upper limits on various species at visible wavelengths
have been presented by {\em Bundy and Marcy} (2000), {\em Moutou et al.\ }(2001, 2003), 
{\em Winn et al.\ }(2004), and {\em Narita et al.\ }(2005).
\\

{\em 3.2.2\ Reflected Light.} Planets shine in reflected light
with a visible-light flux $f_{pl}$ (relative to that of their stars, $f_{\star}$) of
$$
{\left( \frac{f_{pl}}{f_{\star}} \right) }_{\lambda} (\alpha) = \left( \frac{R_{pl}}{a} \right)^{2} p_{\lambda} \, {\Phi}_{\lambda} (\alpha),
$$
where $a$ is the orbital separation, $p_{\lambda}$ is the geometric albedo,
and ${\Phi}_{\lambda} (\alpha)$ is the phase function, which describes the
relative flux at a phase angle $\alpha$ to that at opposition.  Even assuming
an optimistic values for $p_{\lambda}$, hot Jupiters 
present a flux ratio of less than $10^{-4}$ that of their stars.
See {\em Marley et al.\ }(1999), {\em Seager et al.\ }(2000) and {\em Sudarsky et al.\ }(2000)
for theoretical predictions of the reflection spectra and
phase functions of hot Jupiters.

The first attempts to detect this modulation adopted a spectroscopic approach,
whereby a series of spectra spanning key portions of the orbital phase
are searched for the presence of a copy of the stellar spectrum.  For non-transiting
systems, this method is complicated by the need to search over possible values
of the unknown orbital inclination.  The secondary spectrum should be very well
separated spectroscopically, as the orbital velocities for these hot Jupiters
are typically $100\ {\rm km\, s^{-1}}$, much greater than the typical stellar
line widths of $<15\ {\rm km\, s^{-1}}$.  Since the method requires multiple
high signal-to-noise ratio, high-dispersion spectra, only the brightest systems
have been examined.  A host of upper limits have resulted for several systems
(e.g.\ {\em Charbonneau et al.}, 1999; {\em Collier Cameron et al.}, 2002; 
{\em Leigh et al.}, 2003a, 2003b), typically excluding values of $p_{\lambda} > 0.25$
averaged across visible wavelengths.  These upper limits assume a functional
dependence for $\Phi_{\lambda} (\alpha)$ as well as a gray albedo, i.e.\ that 
the planetary spectrum is a reflected copy of the stellar spectrum.

Space-based platforms afford the opportunity to study the albedo and
phase function in a straightforward fashion 
by seeking the photometric modulation of the system light.
The MOST satellite ({\em Walker et al.}, 2003) should be able to
detect the reflected light from several hot Jupiters ({\em Green et al.}, 2003),
or yield upper limits that will severely constrain the atmospheric
models, and campaigns on several systems are completed 
or planned.  The upcoming {\em Kepler Mission} ({\em Borucki et al.}, 2003) 
will search for this effect, and
should identify $100-760$ non-transiting hot Jupiters with orbital periods of 
$P < 7$~d ({\em Jenkins and Doyle}, 2003).
\\

{\em 3.2.3.\ Infrared Emission.} At infrared wavelengths,
the secondary eclipse (i.e.\ the
decrement in the system flux due to the passage of the planet behind the star) permits a 
determination of the planet-to-star brightness ratio.  Since the underlying stellar spectrum
may be reliably assumed from stellar models (e.g.\ {\em Kurucz}, 1992),
such estimates afford the first direct constraints on the emitted spectra of planets orbiting
other Sun-like stars.  In the Rayleigh-Jeans limit, the ratio of
the planetary flux to that of the star is 
$$
{\left( \frac{f_{pl}}{f_{\star}} \right)} \simeq \frac{T_{{\rm eq},pl}}{T_{{\rm eff}, \star}}\,  
{\left( \frac{R_{pl}}{R_{\star}} \right)}^{2}.
$$
The last factor is simply the transit depth.  From Table~1, we can see that the
typical ratio of stellar to planetary temperatures is $3.5-5.5$, leading to
predicted secondary eclipse amplitudes of several millimagnitudes.

{\em Charbonneau et al.\ }(2005) and {\em Deming et al.\ }(2005a) have recently
employed the remarkable sensitivity and stability of the {\em Spitzer Space Telescope}
to detect the thermal emission from TrES-1 (4.5~$\mu$m and 8.0~$\mu$m) 
and HD~209458b (24~$\mu$m); Fig.~4.  These measurements provide estimates of
the planetary brightness temperatures in these 3 bands, which in turn can
be used to estimate (under several assumptions) the value of $T_{{\rm eq},pl}$
and $A$ of the planets.  Observations of these two objects in the other
{\em Spitzer} bands shown in Fig.~4 (as well as the 16~$\mu$m photometric
band of the IRS peak-up array) are feasible.  Indeed, at the time of
writing, partial datasets have been gathered for all four planets shown in Fig.~4.
The results should permit a detailed search for the presence of
spectroscopically-dominant molecules, notably, ${\rm CH_4}$, CO, and ${\rm H_{2}O}$.
Using the related technique of occultation spectroscopy,
{\em Richardson et al.\ }(2003a, 2003b) have analyzed a series of 
infrared spectra spanning a time before, during, and after secondary eclipse, 
and present useful upper limits on the presence of planetary features due to
these molecules.

{\em Williams et al.\ }(2006) have outlined a technique by which the spatial
dependence of the planetary emission could be resolved in longitude through
a careful monitoring of the secondary eclipse. Such observations, as well
as attempts to measure the phase variation as the planet orbits the star
(and hence presents a different face to the Earth) are eagerly anticipated
to address numerous models of the dynamics and weather of these atmospheres
(\S3.1.2).

The elapsed time between the primary and secondary eclipse affords a stringent 
upper limit on the quantity $e \cos{\varpi}$, and the relative durations
of the two events constrains $e \sin{\varpi}$ 
({\em Kallrath and Milone} 1999; {\em Charbonneau} 2003).  The resulting limits on $e$
are of great interest in gauging whether tidal circularization is
a significant source of energy for the planet (\S2.2.4).
\\

{\em 3.2.4.\ Inferences from the Infrared Detections.} Varying planet mass, planet radius,
and stellar mass within their error bars alters the
resulting predicted average planet-star flux ratios only slightly ({\em Burrows et al.}, 2005).
Similarly, and perhaps surprisingly,
adding Fe and forsterite clouds does not shift the predictions in the
Spitzer bands by an appreciable amount.  Moreover, despite
the more than a factor of two difference in the stellar flux at the planet, 
the predictions for the planet-star ratios for TrES-1 and HD209458b are not
very different.
{\em Fortney et al.\ }(2005b) explore the effect of increasing the metallicity
of the planets, and find a better agreement to the red 4.5/8.0~\mic color
of TrES-1 with a enrichment factor of $3-5$.  {\em Seager et al.\ }(2005)
show that models with an increased carbon-to-oxygen abundance produce
good fits to the HD~209458b data, but conclude that a wide range of models produce
plausible fits.  {\em Barman et al.\ }(2005) examine the effect of varying
the efficiency for the redistribution of heat from the dayside to the nightside,
and find evidence that models with significant redistribution (and hence more
isotropic temperatures) are favored.

In Fig.~4, there is a hint
of the presence of H$_2$O, since it is expected to suppress flux between $4-10$~\mic. 
This is shortward of the predicted 10~\mic peak in planet-star flux ratio,
which is due to water's relative abundance and the strength of its
absorption bands in that wavelength range.
Without H$_2$O, the fluxes in the $3.6-8.0$~\mic bands would be much greater.
Hence, a comparison of the TrES-1 and HD~209458b data 
suggests, but does not prove, the presence of water.
Seeing (or excluding) the expected slope between the 5.8~\mic
and 8.0~\mic bands and the rise from 4.5~\mic to 3.6~\mic would be more
revealing in this regard. Furthermore, the relative strength of the 24~\mic flux ratio in
comparison with the 3.6~\mic, 4.5~\mic, and 5.8~\mic
channel ratios is another constraint on the models, as is the
closeness of the 8.0~\mic and 24~\mic ratios. 
If CH$_4$ is present in abundance, then the 3.6~\mic band will test this.  However,
the preliminary conclusion for these close-in Jupiters is that CH$_4$  
should not be in evidence.
Models have difficulty fitting the precise depth of the 4.5~\mic feature for TrES-1.
It coincides with the strong CO
absorption predicted to be a signature of hot Jupiter atmospheres.
However, the depth of this feature is only a weak function of the CO abundance.
A CO abundance 100$\times$ larger than expected in 
chemical equilibrium lowers this flux ratio
at 4.5~\mic by only $\sim$25\%. Therefore, while the 4.5~\mic data
point for TrES-1 implies that CO has been detected, the exact fit is problematic.

In sum, {\em Spitzer} observations of the secondary eclipses of the 
close-in transiting giant planets 
will provide information on the presence of CO and H$_2$O in their
atmospheres, as well as on the role of clouds in 
modifying the planet-to-star flux ratios over the $3-25$~\mic 
spectral range.  Furthermore, there is good reason to believe that 
the surface elemental abundances of extrasolar giant planets are 
not the same as the corresponding stellar elemental abundances, 
and {\em Spitzer} data across the available bandpasses will soon better constrain 
the atmospheric metallicities and C/O ratios of these planets. Moreover, and 
most importantly, the degree to which the heat deposited
by the star on the day side is advected by winds and jet streams to
the night side is unknown.  If this transport is efficient, the day-side
emissions probed during secondary eclipse will be lower than the case
for inefficient transport.  There is already indication in the
data for HD~209458b and TrES-1 that
such transport may be efficient (e.g.\ {\em Barman et al.}, 2005), but much more data are needed to
disentangle the effects of the day-night heat redistribution, metallicity,
and clouds and to identify the diagnostic signatures of the climate 
of these extrasolar giant planets.  The recently detected hot Jupiter, HD~189733b
({\em Bouchy et al.}, 2005a) is a veritable goldmine for such observations (Fig.~4),
owing to the much greater planet-to-star contrast ratio.

\section{\textbf{FUTURE PROSPECTS}}
With the recent radial-velocity discoveries of 
planets with masses of $7-20\ M_{\oplus}$ (e.g.\ {\em Bonfils et al.} 2005;
{\em Butler et al.}, 2004; {\em McArthur et al.}, 2004; {\em Rivera et al.}, 2005;
{\em Santos et al.}, 2004), the identification
of the first such object in a transiting configuration is eagerly awaited.
The majority of these objects have been found in orbit around
low-mass stars, likely reflecting the increased facility of their
detection for a fixed Doppler precision.  Despite the smaller
expected planetary size, the technical challenge of measuring
the transits will be alleviated by the smaller stellar
radius, which will serve to make the transits deep (but less
likely to occur).  Due to the low planetary mass, 
the influence of a central core (\S2.2.2) will be much more prominent.  
Furthermore, the reduced stellar size and brightness
implies that atmospheric observations (\S3.2) will be feasible.
The radial-velocity surveys monitor few stars later than M4V, but
transiting planets of even later spectral types could be identified by
a dedicated photometric monitoring campaign of several thousand
of the nearest targets.  An Earth-sized planet orbiting a 
late M-dwarf with a week-long period would lie
within the habitable zone and, moreover, it would present the same
infrared planet-to-star brightness ratio as that detected (\S3.2.3).  We note the 
urgency of locating  such objects (should they exist), due to the limited cryogenic lifetime 
of {\em Spitzer}.

The excitement with which we anticipate the results from the
{\em Kepler} ({\em Borucki et al.}, 2003) and {\em COROT} ({\em Baglin}, 2003) missions
cannot be overstated.  These projects aim to detect scores of rocky
planets transiting Sun-like primaries, and the {\em Kepler Mission} in particular
will be sensitive to year-long periods and hence true analogs of the Earth.  
Although direct follow-up of such systems (\S3.2) with
extant facilities appears precluded by signal-to-noise considerations,
future facilities (notably the {\em James Webb Space Telescope})
may permit some initial successes.

We conclude that the near-future prospects for studies of transiting
planets are quite bright (although they may dim, periodically), and we anticipate
that the current rapid pace of results will soon eclipse this review --
just in time for Protostars and Planets VI.

\bigskip
\textbf{ Acknowledgments.}
AB acknowledges support from NASA through grant NNG04GL22G.
GL acknowledges support from the NASA OSS and NSF CAREER programs.
We thank Scott Gaudi for illuminating suggestions.

\bigskip

\centerline{\textbf{REFERENCES}}
\bigskip
\parskip=0pt
{\small
\baselineskip=11pt

\refs Agol E., Steffen J., Sari R., and Clarkson W. (2005) \mnras, {\em 359}, 567-579.

\refs Allard F., Hauschildt P.~H., Alexander D.~R., and Starrfield, S. (1997) 
{\em Ann. Rev. Astron. Astrophys., 35}, 137-177.

\refs Allard F., Baraffe I., Chabrier G., Barman T.~S., and Hauschildt P.~H.  (2003)
In {\em Scientific Frontiers in Research on Extrasolar Planets} (D. Deming and S. Seager, eds.), 
pp. 483-490. ASP Conf. Series, San Franciso.

\refs Alonso R. (2005) {\em Ph.D. Thesis}, University of La Laguna.

\refs Alonso R., Brown T.~M., Torres, G., Latham D.~W., Sozzetti A., et al. 
(2004) \apj, {\em 613}, L153-156.

\refs Baglin A. (2003) {\em Adv. Space Res., 31}, 345-349.

\refs Bakos G., Noyes R.~W., Kov{\'a}cs G., Stanek K.~Z., Sasselov D.~D., et al.
(2004) \pasp, {\em 116}, 266-277. 

\refs Baraffe I., Selsis F., Chabrier G., Barman T.~S., Allard F., et al. (2004)
\aap, {\em 419}, L13-L16. 

\refs Barman T.~S., Hauschildt P.~H., and Allard F. (2001) \apj, {\em 556}, 885-895. 

\refs Barman T.~S., Hauschildt P.~H., Schweitzer A., Stancil P.~C., Baron E., et al.
(2002), \apj, {\em 569}, L51-L54.

\refs Barman T.~S., Hauschildt P.~H., and Allard F. (2005), \apj, {\em 632}, 1132-1139. 

\refs Beck J. G. and Giles P. (2005) \apj, {\em 621}, L153-L156.

\refs Bodenheimer P., Lin D.~N.~C., and Mardling R.~A. (2001) \apj, {\em 548}, 466-472.

\refs Bodenheimer P., Laughlin G., and Lin D.~N.~C. (2003) \apj, {\em 592}, 555-563.

\refs Bonev G.~V., Militzer B., and Galli G. (2004) {\em Phys. Rev. B, 69}, 014101.

\refs Bonfils X., Forveille T., Delfosse X., Udry S., Mayor M., et al. (2005) 
\aap {\em 443}, L15-L18.

\refs Boriskov G. V. et al. (2003) {\em Dokl. Phys. 48}, 553-555.

\refs Borucki W.~J., Caldwell D., Koch  D.~G., Webster L.~D., Jenkins J.~M., 
et al. (2001) \pasp, {\em 113}, 439-451. 

\refs Borucki W.~J., Koch D.~G., Lisssauer J.~J., Basri G.~B., 
Caldwell J.~F., et al. (2003) {\em Proc. SPIE, 4854}, 129-140.

\refs Boss A. P. (1997) {\em Science, 276}, 1836-1839.

\refs Boss A. P. (1998) \apj, {\em 503}, 923-937.

\refs Boss A.~P. (2000) \apj, {\em 536}, L101-L104. 

\refs Boss A. P. (2004) \apj, {\em 610}, 456-463.

\refs Bouchy F., Pont F., Santos N.~C., Melo C., Mayor M., et al. (2004)
\aap, {\em 421}, L13-L16.

\refs Bouchy F., Udry S., Mayor M., Pont F., Iribane N., et al. (2005a)
\aap, {\em 444}, L15-L19.

\refs Bouchy F., Pont F., Melo C., Santos N.~C., Mayor M., et al. (2005b)
\aap, {\em 431}, 1105-1121.

\refs Brown T.~M. (2001), \apj, {\em 553}, 1006-1026. 

\refs Brown T.~M. (2003) \apj, {\em 593}, L125-L128.

\refs Brown T.~M. and Charbonneau D. (2000).
In {\em Disks, Planetesimals, and Planets} (F. Garz{\'o}n et al., eds.),
pp. 584-589. ASP Conf. Series, San Francisco.

\refs Brown T.~M., Charbonneau D., Gilliland R.~L., Noyes R.~W., and Burrows A. 
(2001) \apj, {\em 551}, 699-709.

\refs Brown T.~M., Libbrecht K.~G., and Charbonneau D. (2002), \pasp, {\em 114}, 826-832. 
 
\refs Bundy K.~A. and Marcy G.~W.(2000), \pasp {\em 112}, 1421-1425.

\refs Burke C.~J., Gaudi B.~S., DePoy D.~L., Pogge R.~W., and Pinsonneault M.~H. 
(2004), \aj, {\em 127}, 2382-2397. 

\refs Burkert A., Lin D.~N.~C., Bodenheimer P.~H., Jones C.~A., and Yorke H.~W.
(2005) \apj, {\em 618}, 512-523.

\refs Burrows A. (2005) {\em Nature, 433}, 261-268.

\refs Burrows A., Marley M., Hubbard W.~B., Lunine J.~I., Guillot T., et al. (1997)
\apj, {\em 491}, 856-875.

\refs Burrows A., Hubbard W.~B., Lunine J.~I., and Liebert J. (2001)
{\em Rev. Mod. Phys., 73}, 719-765.

\refs Burrows A., Sudarsky D., and Hubbard W.~B. (2003) \apj, {\em 594}, 545-551.

\refs Burrows A., Hubeny I., Hubbard W.~B., Sudarsky D., and Fortney J.~J. (2004)
\apj, {\em 610}, L53-L56.

\refs Burrows A., Sudarsky D., and Hubeny I. (2004) \apj, {\em 609}, 407-416.

\refs Burrows A., Hubeny I., and Sudarsky D. (2005) \apj, {\em 625}, L135-L138.

\refs Butler R.~P., Vogt S.~S., Marcy  G.~W., Fischer D.~A., 
Wright J.~T., et al. (2004) \apj, {\em 617}, 580-588. 

\refs Chabrier G., Barman T., Baraffe I., Allard F., and Hauschildt P.~H.
(2004) \apj, {\em 603}, L53-L56.

\refs Charbonneau D. (2003)
In {\em Scientific Frontiers in Research on Extrasolar Planets} (D. Deming and S. Seager, eds.), 
pp. 449-456. ASP Conf. Series, San Franciso.

\refs Charbonneau D., Noyes R.~W., Korzennik S.~G., Nisenson P., Jha S., et al. (1999)
\apj, {\em 522}, L145-L148.

\refs Charbonneau D., Brown T.~M., Latham D.~W., and Mayor M. (2000)
\apj, {\em 529}, L45-L48.

\refs Charbonneau D., Brown T.~M., Noyes R.~W., and Gilliland R.~L. (2002)
\apj, {\em 568}, 377-384.

\refs Charbonneau D., Brown T.~M., Dunham E.~W., Latham D.~W., Looper D.~L., et al. (2004)
In {\em The Search for Other Worlds} (S. Holt and D. Deming, eds.), pp. 151-160. 
AIP Conf. Series.
 
\refs Charbonneau D., Allen, L.~E., Megeath S.~T., Torres G., Alonso R.,
et al. (2005) \apj, {\em 626}, 523-529.

\refs Charbonneau D., Winn J.~N., Latham D.~W., Bakos G., Falco E., et al.
(2006), \apj, {\em 636}, 445-452.

\refs Cho J.~Y.-K., Menou K., Hansen B.~M.~S., and Seager S. (2003) \apj, {\em 587},
L117-L120.

\refs Christian D.~J., Pollacco D.~L., Clarkson W.~I., Collier Cameron A., Evans N., et al.
(2004) In {\em The 13th Cool Stars Workshop} (F. Favata, ed.).
ESA Spec. Pub. Series.

\refs Cody A.~M. and Sasselov D.~D. (2002) \apj, {\em 569}, 451-458.

\refs Collier Cameron A., Horne K., Penny A., and Leigh C. (2002), \mnras, {\em 330}, 187-204. 
 
\refs Collins G. W., d Silva L. B., Celliers P., et al. (1998) {\em Science, 281}, 1178-1181.

\refs Cooper C.~S. and Showman A.~P. (2005) \apj, {\em 629}, L45-L48.

\refs da Silva R., Udry S., Bouchy F., Mayor M., Moutou C., et al. (2006) \aap, {\em 446}, 717-722.

\refs Deeg H.~J., Garrido R., and Claret A. (2001) {\em New Astronomy, 6}, 51-60. 

\refs Deeg H.~J., Alonso R., Belmonte J.~A., Alsubai K., Horne K., et al. (2004)
\pasp, {\em 116}, 985-995. 

\refs Deming D., Seager S., Richardson L.~J., and Harrington J. (2005a) {\em Nature, 434}, 740-743.

\refs Deming D., Brown T.~M., Charbonneau D., Harrington J., and Richardson L.~J. (2005b), 
\apj, {\em 622}, 1149-1159. 

\refs Desjarlais M.~P. (2003) {\em Phys Rev. B, 68}, 064204.

\refs Drake A.~J. (2003) \apj, {\em 589}, 1020-1026.

\refs Dreizler S., Rauch T., Hauschildt P., Schuh S.~L., Kley W., et al. (2002)
\aap, {\em 391}, L17-L20. 

\refs Dunham E.~W., Mandushev G.~I., Taylor B.~W., and Oetiker B. (2004)
\pasp, {\em 116}, 1072-1080. 

\refs Fischer D., Laughlin G., Butler R.~P., Marcy G., Johnson J., et al. (2005)
\apj, {\em 620}, 481-486.

\refs Fortney, J.~J. (2005) \mnras, {\em 364}, 649-653.

\refs Fortney J.~J. and Hubbard W.~B. (2003) {\em Icarus, 164}, 228-243.

\refs Fortney J.~J., Sudarsky D., Hubeny I., Cooper C.~S., Hubbard W.~B., et al. (2003)
\apj, {\em 589}, 615-622.

\refs Fortney J.~J., Marley M.~S., Lodders K., Saumon D., and Freedman R.~S.
(2005a) \apj, {\em 627}, L69-L72.

\refs Fortney J.~J., Saumon D., Marley M.~S., Lodders K., and Freedman R.
(2005b) \apj, in press.

\refs Gaudi B.~S. (2005) \apj, {\em 628}, L73-L76.

\refs Gaudi B.~S., Seager S., and Mallen-Ornelas G. (2005) \apj, {\em 623}, 472-481.

\refs Gilliland R.~L., Brown T.~M., Guhathakurta P., Sarajedini A., Milone E.~F., et al. (2000)
\apj, {\em 545}, L47-L51.

\refs Girardi L., Bertelli G., Bressan A., Chiosi C., Groenewegen M.~A.~T., et al. (2002)
\aap, {\em 391}, 195-212.

\refs Gould A., Pepper J., and DePoy D.~L. (2003) \apj, {\em 594}, 533-537.

\refs Green D., Matthews J., Seager S., and Kuschnig R. (2003) \apj, {\em 597}, 590-601.

\refs Greenberg R. (1974) {\em Icarus, 23}, 51-58. 

\refs Guillot T. (2005) {\em Ann. Rev. Earth Planet. Sci, 33}, 493-530.

\refs Guillot T. and Showman A.~P. (2002) \aap, {\em 385}, 156-165.

\refs Guillot T., Burrows A., Hubbard W.~B., Lunine J.~I., and Saumon D. (1996)
\apj, {\em 459}, L35-L38.

\refs Hartman J.~D., Stanek K.~Z., Gaudi B.~S., Holman M.~J., and McLeod B.~A. (2005)
\aj, {\em 130}, 2241-2251.

\refs Henry G.~W., Marcy G.~W., Butler R.~P., and Vogt S.~S. (2000)
\apj, {\em 529}, L41-L44.

\refs Holman M.~J. and Murray N.~W. (2005) {\em Science, 307}, 1288-1291.

\refs Holman M.~J., Winn J.~N., Stanek K.~Z., Torres G., Sasselov D.~D., et al. (2005)
\apj, submitted.

\refs Holmes N.~C., Ross M, and Nellis W.~J. (1995) {\em Phys. Rev. B., 52}, 15835-15845.

\refs Hubbard W.~B. (1968) \apj, {\em 152}, 745-754.

\refs Hubbard W.~B. (1977) {\em Icarus, 30}, 305-310.

\refs Hubbard W.~B., Fortney J.~J., Lunine J.~I., Burrows A., Sudarsky D., et al. (2001)
\apj, {\em 560}, 413-419.

\refs Hubickyj O., Bodenheimer P., and Lissauer J.~J. (2004)
In {\em Gravitational Collapse: From Massive Stars to Planets}
(G. Garc{\'{\i}}a-Segura et al., eds), pp. 83-86. 
Rev. Mex. Astron. Astrophys. Conf. Series.

\refs Hut P. (1980) \aap, {\em 92}, 167-170. 

\refs Janes K. (1996) {\em J. Geophys. Res.}, {\em 101}, 14853-14860. 

\refs Jenkins J.~M. and Doyle L.~R. (2003) \apj, {\em 595}, 429-445. 

\refs Jenkins J.~M., Caldwell D.~A., and Borucki W.~J. (2002) \apj, {\em 564}, 495-507.

\refs Jha S., Charbonneau D., Garnavich P.~M., Sullivan D.~J., Sullivan T., et al. (2000)
\apj, {\em 540}, L45-L48.

\refs Kallrath J. and Milone E.~F. (1999) 
In {\em Eclipsing Binary Stars: Modeling and Analysis} pp. 60-64. 
Springer, New York.

\refs Knudson M.~D., Hanson D.~L., Bailey J.~E., Hall C.~A., Asay J.R., et al. (2004)
{\em Phys. Rev. B, 69}, 144209.

\refs Knutson H., Charbonneau D., Noyes R.~W., Brown T.~M., and Gilliland R.~L. (2006)
\apj, submitted.

\refs Konacki M., Torres G., Jha S., and Sasselov D.~D. (2003a) {\em Nature, 421}, 507-509.

\refs Konacki M., Torres G., Sasselov D.~D., and Jha S. (2003b) \apj, {\em 597}, 1076-1091. 

\refs Konacki M., Torres G., Sasselov D.~D., Pietrzynski G., Udalski A., et al. (2004)
\apj, {\em 609}, L37-L40.

\refs Konacki M., Torres G., Sasselov D.~D., and Jha S. (2005)
\apj, {\em 624}, 372-377.

\refs Kov{\'a}cs G., Zucker S., and Mazeh T. (2002), \aap, {\em 391}, 369-377. 

\refs Kurucz R. (1992)
In {\em The Stellar Populations of Galaxies} (B. Barbuy and A. Renzini, eds.), pp. 225-232.
Kluwer, Dordrecht.

\refs Latham D.~W. (2003)
In {\em Scientific Frontiers in Research on Extrasolar Planets} (D. Deming and S. Seager, eds.), 
pp. 409-412. ASP Conf. Series, San Franciso.

\refs Laughlin G., Wolf A., Vanmunster T., Bodenheimer P., Fischer D., et al. (2005a)
\apj, {\em 621}, 1072-1078.

\refs Laughlin G., Marcy G.~W., Vogt, S.~S., Fischer D.~A., and Butler R.~P. (2005b) 
\apj, {\em 629}, L121-L124. 

\refs Lecavelier des Etangs A., Vidal-Madjar A., McConnell J.~C., and H{\'e}brard G. (2004)
\aap, {\em 418}, L1-L4. 

\refs Leigh C., Collier Cameron A., Udry S., Donati J.-F., Horne K., et al. (2003a)
\mnras, {\em 346}, L16-L20. 
 
\refs Leigh C., Collier Cameron A., Horne K., Penny A., and James D. (2003b)
\mnras, {\em 344}, 1271-1282.

\refs Lin D.~N.~C., Bodenheimer P., and Richardson D.~C. (1996) {\em Nature, 380}, 606-607. 

\refs Lindal G.~F., Wood G.~E., Levy G.~S., Anderson J.~D., Sweetnam D.~N., et al. (1981)
{\em \jgr}, {\em 86}, 8721-8727.

\refs Lissauer J.~J. (1993) {\em Ann. Rev. Astron. Astrophys., 31}, 129-174. 

\refs Mallen-Ornelas G., Seager S., Yee H.~K.~C., Minniti D., Gladders M.~D. et al. (2003)
\apj, {\em 582}, 1123-1140.

\refs Mandel K. and Agol E. (2002) \apj, {\em 580}, L171-L174.

\refs Mandushev G., Torres G., Latham D.~W., Charbonneau D., Alonso R., et al. (2005) \apj,
{\em 621}, 1061-1071.

\refs Marley M.~S., Gelino C., Stephens D., Lunine J.~I., and Freedman R. (1999)
\apj, {\em 513}, 879-893. 

\refs Marshall J.~L., Burke C.~J., DePoy D.~L., Gould A., and Kollmeier J.~A. (2005)
\aj, {\em 130}, 1916-1928. 

\refs Mayor M. and Queloz D. (1995) {\em Nature, 378}, 355-359.

\refs Mazeh T., Naef D., Torres G., Latham D.~W., Mayor M., et al. (2000)
\apj, {\em 532}, L55-L58.

\refs McArthur B.~E., Endl M., Cochran W.~D., Benedict G.~F., Fischer D.~A., et al. (2004)
\apj {\em 614}, L81-L84. 

\refs McCullough P.~R., Stys J.~E., Valenti J.~A., Fleming S.~W., Janes K.~A., et al. (2005)
\pasp, {\em 117}, 783-795. 

\refs McLaughlin D.~B. (1924) \apj, {\em 60}, 22-31.

\refs Menou K, Cho J.~Y-K., Hansen B.~M.~S., and Seager S. (2003) \apj, {\em 587}, L113-L116.

\refs Militzer B. and Ceperley D.~M. (2001) {\em Phys. Rev. E, 63}, 066404.

\refs Miralda-Escud{\'e} J. (2002) \apj, {\em 564}, 1019-1023. 

\refs Mizuno H. (1980) {\em Prog. Theor. Phys., 64}, 544-557.
 
\refs Mochejska B.~J., Stanek K.~Z., Sasselov D.~D., Szentgyorgyi A.~H.,
Bakos G.~{\'A}., et al. (2005) \aj, {\em 129}, 2856-2868. 
 
\refs Mochejska B.~J., Stanek K.~Z., Sasselov D.~D., Szentgyorgyi A.~H.,
Adams E. et al. (2006) \aj, {\em 131}, 1090-1105. 

\refs Moutou C., Coustenis A., Schneider J., St Gilles R., Mayor M., et al. (2001)
\aap, {\em 371}, 260-266. 

\refs Moutou C., Coustenis A., Schneider J., Queloz D., and Mayor M. (2003)
\aap, {\em 405}, 341-348. 

\refs Moutou C., Pont F., Bouchy F., and Mayor M. (2004) \aap, {\em 424}, L31-L34.

\refs Murray N., Hansen B., Holman M, and Tremaine S. (1998), {\em Science, 279}, 69-72.

\refs Narita N., Suto Y., Winn J.~N., Turner E.~L., Aoki W., et al. (2005)
\pasj, {\em 57}, 471-480.

\refs O'Donovan, F.~T., Charbonneau D., Torres G., Mandushev G., Dunham E.~W., et al. (2006)
\apj, in press.

\refs Ohta Y., Taruya A., and Suto Y. (2005) \apj, {\em 622}, 1118-1135.

\refs Peale S.~J. (1969) \aj, {\em 74}, 483-489.

\refs Pepper J. and Gaudi B.~S. (2005) \apj, {\em 631}, 581-596. 

\refs Pepper J. and Gaudi B.~S. (2006) {\em Acta Astron.}, submitted.

\refs Pepper J., Gould A., and Depoy, D.~L. (2003) {\em Acta Astron.}, {\em 53}, 213-228. 

\refs Pepper J., Gould A., and Depoy D.~L. (2004)
In {\em The Search for Other Worlds} (S. Holt and D. Deming, eds.), pp. 185-188. 
AIP Conf. Series.

\refs Pollack J.~B. Hubickyj O., Bodenheimer P., Lissauer J.~J., Podolak M., et al. (1996)
{\em Icarus, 124} 62-85.

\refs Pont F., Bouchy F., Queloz D., Santos N.~C., Melo C., et al. (2004) \aap, {\em 426}, L15-L18.

\refs Pont F., Bouchy F., Melo C., Santos N.C., Mayor M., et al. (2005) \aap, {\em 438}, 1123-1140. 

\refs Queloz D., Eggenberger A., Mayor M., Perrier C., Beuzit J.~L., et al. (2000)
\aap, {\em 359}, L13-L17. 

\refs Rauer H., Eisl{\"o}ffel J., Erikson A., Guenther E., Hatzes A.~P., et al. (2004)
\pasp, {\em 116}, 38-45. 

\refs Richardson L.~J., Deming D., and Seager S. (2003a) \apj, {\em 597}, 581-589. 
 
\refs Richardson L.~J., Deming D., Wiedemann G., Goukenleuque C., Steyert D., et al. (2003b)
\apj, {\em 584}, 1053-1062.

\refs Rivera E.~J., Lissauer J.~J., Butler R.~P., Marcy G.~W., Vogt S.~S., et al. (2005)
\apj, {\em 634}, 625-640. 

\refs Rossiter R.~A. (1924) \apj, {\em 60}, 15-21.

\refs Santos N.~C., Bouchy F., Mayor M., Pepe F., Queloz D., et al. (2004)
\aap, {\em 426}, L19-L23. 

\refs Sato B., Fischer D.~A., Henry G.~W., Laughlin G., Butler R.~P., et al. (2005)
\apj, {\em 633}, 465-473. 

\refs Saumon D. and Guillot T. (2004) \apj, {\em 460}, 993-1018.

\refs Saumon D., Chabrier G., Wagner D.~J., and Xie X. (2000) {\em High Press. Res., 16}, 331-343.

\refs Seager S. and Mallen-Ornelas G. (2003) \apj, {\em 585}, 1038-1055.

\refs Seager S. and Sasselov D.~D. (1998) \apj, {\em 502}, L157-L161.

\refs Seager S. and Sasselov D.~D. (2000) \apj, {\em 537}, 916-921. 

\refs Seager S., Whitney B.~A, and Sasselov D.~D. (2000) \apj, {\em 540}, 504-520.

\refs Seager S., Richardson L.~J., Hansen B.~M.~S., Menou K., Cho J.~Y-K., et al. (2005)
\apj, {\em 632}, 1122-1131.

\refs Showman A.~P. and Guillot T. (2002) \aap, {\em 385}, 166-180.

\refs Silva A.~V.~R. (2003) \apj, {\em 585}, L147-L150. 

\refs Sirko E. and Paczy\'nski B. (2003) \apj, {\em 592}, 1217-1224.

\refs Sozzetti A., Yong D., Torres G., Charbonneau D., Latham D.~W., et al. (2004)
\apj, {\em 616}, L167-L170.

\refs Steffen J.~H. and Agol E. (2005), \mnras {\em 364}, L96-L100. 

\refs Struve O. (1952) {\em Observatory, 72}, 199-200.

\refs Sudarsky D., Burrows A., and Pinto P. (2000) \apj, {\em 538}, 885-903. 
 
\refs Sudarsky D., Burrows A., and Hubeny I. (2003) \apj, {\em 588}, 1121-1148.

\refs Tingley B. (2004) \aap, {\em 425}, 1125-1131.

\refs Torres G., Konacki M., Sasselov D.~D., and Jha S. (2004a), \apj, {\em 609}, 1071-1075.

\refs Torres G., Konacki M., Sasselov D.~D., and Jha S. (2004b), \apj, {\em 614}, 979-989.

\refs Torres G., Konacki M., Sasselov D.~D., and Jha S. (2005), \apj, {\em 619}, 558-569.

\refs Udalski A., Paczynski B., Zebrun K., Szymaski M., Kubiak M., et al. (2002a)
{\em Acta Astron., 52}, 1-37.

\refs Udalski A., Zebrun K., Szymanski M., Kubiak M., Soszynski I., et al. (2002b)
{\em Acta Astron., 52}, 115-128.

\refs Udalski A., Szewczyk O., Zebrun K., Pietrzynski G., Szymanski M., et al. (2002c)
{\em Acta Astron., 52}, 317-359.

\refs Udalski A., Pietrzynski G., Szymanski M., Kubiak M., Zebrun K., et al. (2003)
{\em Acta Astron., 53}, 133-149.

\refs Udalski A. Szymanski M.~K., Kubiak M., Pietrzynski G., Soszynski I., et al. (2004)
{\em Acta Astron., 54}, 313-345.

\refs Vidal-Madjar A., Lecavelier des Etangs A.,
D{\'e}sert J.-M., Ballester G.~E., Ferlet R., et al. (2003) Nature, {\em 422}, 143-146.

\refs Vidal-Madjar A., D{\'e}sert J.-M., Lecavelier des Etangs A.,
H{\'e}brard G., Ballester G.~E., et al. (2004) \apj, {\em 604}, L69-L72.

\refs von Braun K., Lee B.~L., Seager S., Yee H.~K.~C., Mall{\'e}n-Ornelas G., et al.
(2005) \pasp, {\em 117}, 141-159. 

\refs Walker G., Matthews J., Kuschnig R., Johnson R., Rucinski S., et al. (2003)
\pasp, {\em 115}, 1023-1035.

\refs Williams P.~K.~G., Charbonneau D., Cooper C.~S., Showman A.~P., and Fortney J.~J. (2006)
\apj, submitted.

\refs Winn J. and Holman M.~J. (2005) \apj, {\em 628}, L159-L162.

\refs Winn J.~N., Suto Y., Turner E.~L., Narita N., Frye B.~L., et al. (2004)
\pasj, {\em 56}, 655-662. 

\refs Winn J.~N., Noyes R.~W., Holman M.~J., Charbonneau D., Ohta Y., et al. (2005)
\apj, {\em 631}, 1215-1226.

\refs Wittenmyer R.~A., Welsh W.~F., Orosz J.~A., Schultz A.~B., Kinzel W., et al. (2005)
\apj, {\em 632}, 1157-1167. 

\refs Wolf A., Laughlin G., Henry G.~W., Fischer D.~A., Marcy G., et al. (2006)
\apj, in press.

}

\end{document}